\documentclass[aps,pra,preprintnumbers,showpacs,tightenlines]{revtex4}

\usepackage{amssymb}
\usepackage{amsmath}
\usepackage{graphicx}
\usepackage{epsfig}
\usepackage{subfigure}
\usepackage{amsfonts}
\usepackage{CJK}

\begin{document}

\title{Multiqubit tunable phase gate of one qubit simultaneously
controlling $n$ qubits in a cavity}

\author{Chui-Ping Yang$^{1,2,3,5}$, Shi-Biao Zheng$^{1,4}$, and Franco Nori$^{1,2}$}

\address{$^1$Advanced Science Institute, RIKEN, Wako-Shi, Saitama
351-0198, Japan}

\address{$^2$Physics Department, The University of Michigan,
Ann Arbor, Michigan 48109-1040, USA}

\address{$^3$Department of Physics, Hangzhou Normal University, Hangzhou, Zhejiang 310036, China}

\address{$^4$Department of Physics and State Key Laboratory
Breeding Base of Photocatalysis, Fuzhou University, Fuzhou 350002,
China}

\address{$^5$State Key Laboratory of Precision Spectroscopy, Department of Physics,
East China Normal University, Shanghai 200062, China}

\date{\today}

\begin{abstract}
We propose how to realize a multiqubit tunable phase gate of one
qubit simultaneously controlling $n$ qubits with four-level
quantum systems in a cavity or coupled to a resonator. Each of the
$n$ two-qubit controlled-phase (CP) gates involved in this
multiqubit phase gate has a shared control qubit but a {\it
different} target qubit. In this propose, the two lowest levels of
each system represent the two logical states of a qubit while the
two higher-energy intermediate levels are used for the gate
implementation. The method presented here operates essentially by
creating a single photon through the control qubit, which then
induces a phase shift to the state of each target qubit. The phase
shifts on each target qubit can be adjusted by changing the Rabi
frequencies of the pulses applied to the target qubit systems. The
operation time for the gate implementation is independent of the
number of qubits, and neither adjustment of the qubit level
spacings nor adjustment of the cavity mode frequency during the
gate operation is required by this proposal. It is also noted that
this approach can be applied to implement certain types of
significant multiqubit phase gates, e.g., the multiqubit phase
gate consisting of $n$ two-qubit CP gates which are key elements
in quantum Fourier transforms. A possible physical implementation
of our approach is presented. Our proposal is quite general, and
can be applied to physical systems such as various types of
superconducting devices coupled to a resonator and trapped atoms
in a cavity.
\end{abstract}

 \maketitle
\date{\today}

\begin{center}
\textbf{I. INTRODUCTION}
\end{center}

During the past decade, various physical systems have been considered for
building up quantum information processors [1-4]. Among them, cavity QED
analogs with solid-state systems are particularly appealing [5-10].
Theoretically, it was predicted that the strong coupling limit, which is
difficult to achieve with atoms in a microwave cavity, can be readily
realized with superconducting charge qubits [11-13], superconducting flux
qubits [14], or semiconducting quantum dots [15]. Moreover, the strong
coupling cavity QED has been experimentally demonstrated with
superconducting qubits [16,17] and semiconductor quantum dots embedded in a
microcavity [18]. These experimental results make solid-state qubit cavity
QED a very attractive approach to quantum information processing. The goal
of this work is to try to find ways to implement multiqubit gates in solid
state QED qubit systems. Let us first introduce two-qubit gates and then
generalize to multiqubit gates.

\begin{center}
\textbf{A. Two-qubit controlled phase gates}
\end{center}

It has been shown that standard two-qubit controlled-phase (CP) gates,
together with single-qubit gates, form the building blocks of quantum
information processors. A standard two-qubit CP gate is described by the
transformation $\left| 00\right\rangle \rightarrow \left| 00\right\rangle
,\left| 01\right\rangle \rightarrow \left| 01\right\rangle ,\left|
10\right\rangle \rightarrow \left| 10\right\rangle ,$ and $\left|
11\right\rangle \rightarrow -\left| 11\right\rangle $. This shows that: when
the control qubit (the first qubit) is in the state $\left| 1\right\rangle ,$
the gate flips the phase of the state $\left| 1\right\rangle $ of the target
qubit (the second qubit) by $\pi $ (i.e., $\left| 1\right\rangle \rightarrow
-\left| 1\right\rangle $) but does nothing otherwise. So far, a large number
of theoretical proposals for realizing this two-qubit CP gate have been
presented with many physical systems. Moreover, this two-qubit CP gate,
together with a two-qubit controlled-not (CNOT) gate or a two-qubit $i$SWAP
gate, has been experimentally demonstrated in, e.g., cavity QED [19,20], NMR
[21], and superconducting qubits [22-25].

In this paper, we focus on a different type of two-qubit CP gate. To see the
difference between this type of gate and the standard CP gate above, let us
first consider a two-qubit CP gate with qubit $1$ (the control qubit) and
qubit $k$ (the target qubit). Here, $k$ is the label of the target qubit,
which is an integer greater than 1. The operator describing this gate is
given by the following matrix
\begin{equation}
R_{1k}\left( \theta _k\right) =\left(
\begin{array}{cccc}
1 & 0 & 0 & 0 \\
0 & 1 & 0 & 0 \\
0 & 0 & 1 & 0 \\
0 & 0 & 0 & e^{i\theta _k}
\end{array}
\right) ,
\end{equation}
where $0\leq \theta _k\leq 2\pi $. Namely, if and only if the control qubit $%
1$ is in the state $\left| 1\right\rangle $, a phase shift $e^{i\theta _k}$
happens to the state $\left| 1\right\rangle $ of the target qubit $k.$ It is
obvious that for $\theta _k\neq \pi $, this two-qubit \textit{tunable} CP
gate of Eq.~(1) is different from the standard two-qubit CP gate. It is
called tunable, because $\theta _k$ can be adjusted, as described below.

\begin{figure}[tbp]
\includegraphics[bb=114 382 528 689, width=10.6 cm, clip]{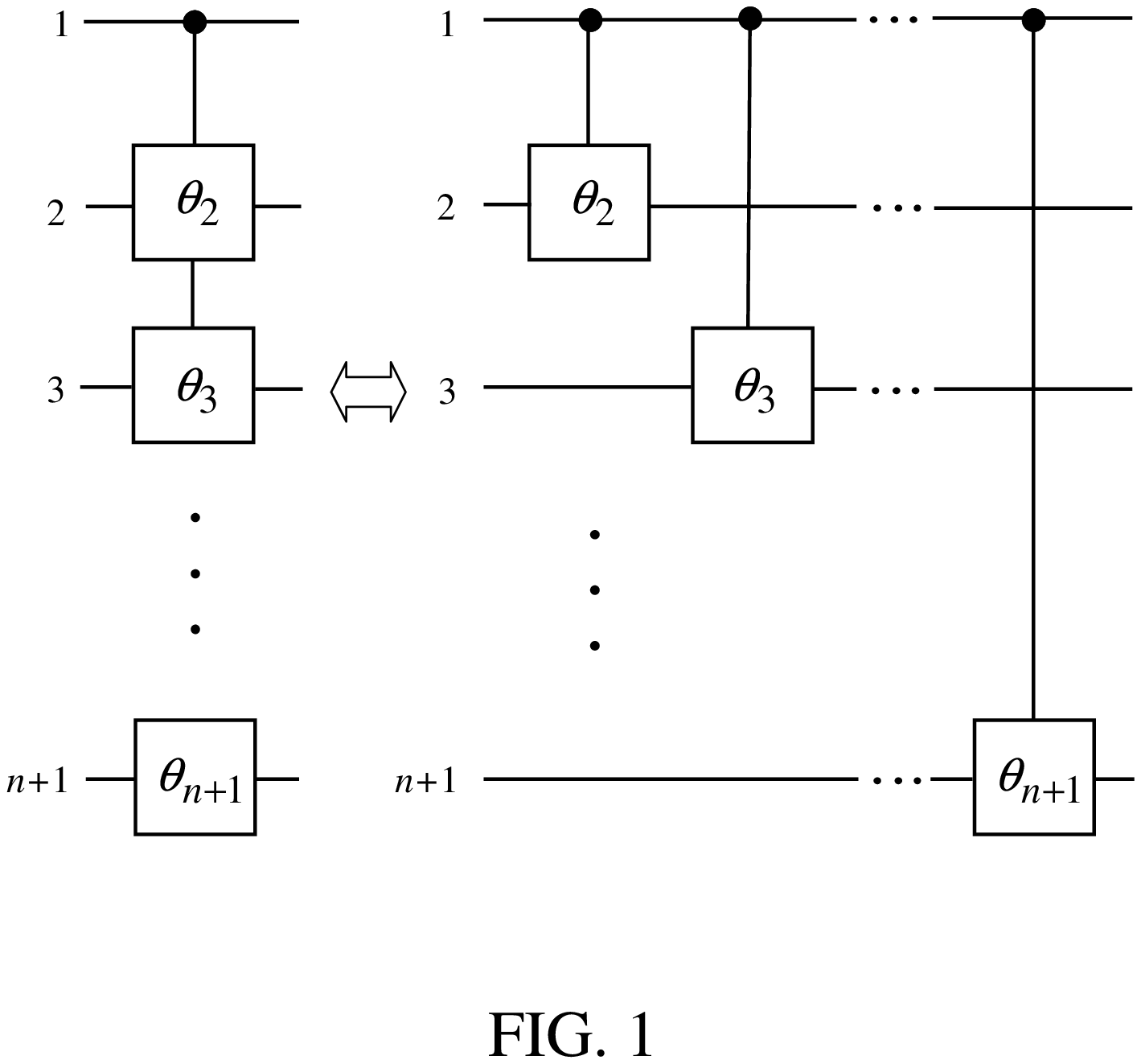} %
\vspace*{-0.08in}
\caption{Schematic circuit of a multiqubit controlled-phase (CP) gate with
one qubit (on top) \textit{simultaneously} controlling $n$ target qubits.
The circuit on the left side is equivalent to the circuit on the right side.
Each of the $n$ two-qubit CP gates forming this multiqubit phase gate has a
shared control qubit (i.e., qubit $1$, represented by the filled circle
located on top) but a different target qubit (i.e., either qubit $2,$ $3,$ $%
...$ $,$ or $n+1$). The element $\theta _k$ represents a controlled-phase
shift by $e^{i\theta _k}$, which is performed on the target qubit $k$ when
the control qubit $1$ is in the state $\left| 1\right\rangle $ ($%
k=2,3,...,n+1$). Namely, if the control qubit $1$ is in the state $\left|
1\right\rangle $, then a phase shift $e^{i\theta _k}$ happens to the state $%
\left| 1\right\rangle $ of the target qubit $k$; otherwise nothing happens.}
\label{fig:1}
\end{figure}

\begin{center}
\textbf{B. Multi-qubit phase gates}
\end{center}

Let us now consider a multiqubit quantum phase gate which consists of $n$
two-qubit CP gates as depicted in Fig. 1. From Fig. 1, one can see that the $%
n$ two-qubit CP gates act on different qubit pairs ($1,2$), ($1,3$), ...,
and ($1,n+1$), respectively. Each two-qubit CP gate has a common control
qubit (qubit $1$) but a \textit{different} target qubit ($2,3,...,$ or $n+1$%
). The CP gate acting on the qubit pair ($1,k$) is described by the operator
$R_{1k}$ above, which induces a phase shift $e^{i\theta _k}$ to the state $%
\left| 1\right\rangle $ of the target qubit $k$ when the control qubit $1$
is in the state $\left| 1\right\rangle .$ In the general case, any two of
the phase shifts $e^{i\theta _2},e^{i\theta _3},...,e^{i\theta _{n+1}}$ for
the $n$ two-qubit CP gates can be the same or different, and any one of them
can be arbitrary. The operator characterizing this multiqubit phase gate
with one qubit simultaneously controlling the $n$ target qubits is

\begin{equation}
U=\otimes _{k=2}^{n+1}R_{1k},
\end{equation}
where the operator $R_{1k}$ is given by Eq.~(1).

\begin{center}
\textbf{C. Motivation for studying multi-qubit gates}
\end{center}

There are several motivations for this work:

(i) Quantum gates with multiple control qubits or multiple target qubits are
of great importance in quantum information processing such as realizing
quantum error-correction protocols, constructing quantum circuits, and
implementing quantum algorithms. When using the conventional
gate-decomposition protocols to construct a multiqubit controlled gate
[26,27], the procedure usually becomes complicated (especially for a large $%
n $), as the number of single-qubit and two-qubit gates required for the
gate implementation heavily depends on the number $n$ of qubits, and
therefore building a multiqubit gate may become very difficult.

(ii) Recently, attention is shifting to the physical realization of
multiqubit gates (e.g., [28]). Although several methods for implementing
multiqubit gates based on cavity QED or trapped ions have been proposed
[29-32], realization of \textit{controllable} multiqubit phase gates based
on cavity QED or ion traps has not been thoroughly investigated.

\begin{figure}[tbp]
\includegraphics[bb=143 345 509 574, width=10.6 cm, clip]{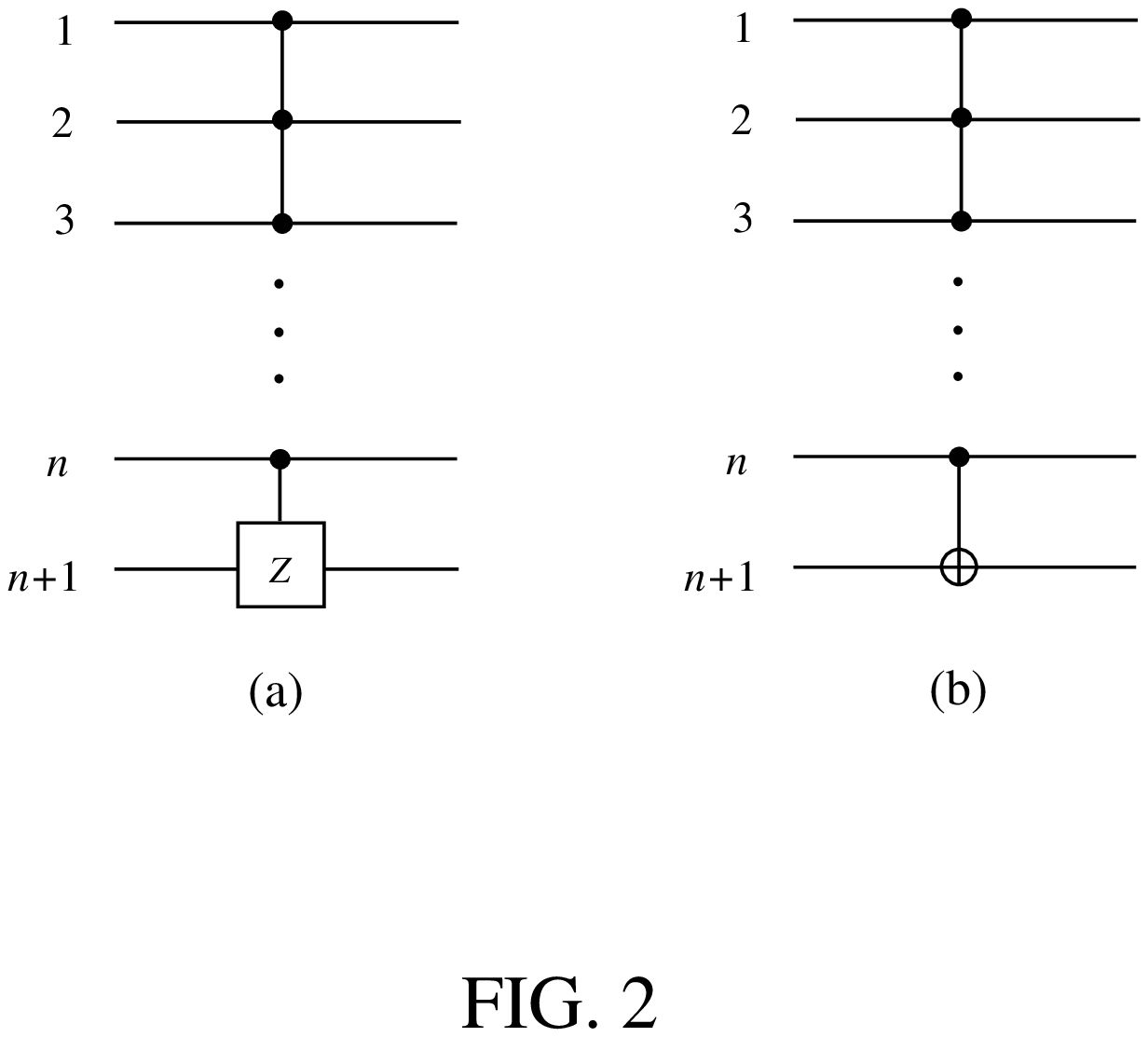} %
\vspace*{-0.08in}
\caption{(a) Schematic circuit of a multiqubit controlled-phase gate with $n$
control qubits acting on one target qubit located at the bottom of the
circuit shown here. The element $Z$ represents a Pauli rotation $\sigma _z$,
i.e., a phase shift by $\pi$ around the $z$-axis (with $n$ control qubits on
the filled circles). Namely, if the $n$ control qubits are \textit{all} in
the state $\left| 1\right\rangle $, then the state $\left| 1\right\rangle $
at $Z$ is phase-shifted by $\pi$ (i.e., $\left| 1\right\rangle \rightarrow
e^{i\pi}\left| 1\right\rangle=-\left| 1\right\rangle $); otherwise nothing
happens to the state at $Z$. (b) Schematic circuit of a multiqubit
controlled-NOT gate with $n$ control qubits acting on one target qubit. The
symbol $\oplus $ represents a controlled-NOT gate (with $n$ control qubits
on the filled circles). If the $n$ control qubits are \textit{all} in the
state $\left| 1\right\rangle $, then the state at $\oplus $ is bit-flipped
(i.e., $\left| 1\right\rangle \rightarrow \left| 0\right\rangle $ and $%
\left| 0\right\rangle \rightarrow \left| 1\right\rangle $).}
\label{fig:2}
\end{figure}

(iii) The methods previously proposed [29-32] are mainly for realizing
quantum controlled-phase or controlled-not gates with \textit{%
multiple-control} qubits acting on \textit{one target} qubit (Fig. 2). These
methods provide a simple and/or fast way for realizing quantum gates with
multiple control qubits (a type of multiqubit gate significant in quantum
information processing), when compared with the conventional
gate-decomposition approaches. However, it is noted that these
proposals~cannot be extended to perform a different kind of important
multiqubit phase gate, i.e., quantum phase gates with one qubit
simultaneously controlling multiple target qubits.

\begin{center}
\textbf{D. Advantages}
\end{center}

The purpose of this work is to find a way of implementing a multiqubit
\textit{tunable} phase gate of one qubit simultaneously controlling $n$
qubits with four-level quantum systems in a cavity or coupled to a
resonator. Our proposal operates essentially by creating a single photon
through the control qubit, which then induces a phase shift to the state of
each target qubit. As shown below, this proposal has the following
advantages:

(i) since the excited level $\left| 3\right\rangle $ is unpopulated during
the gate operation, decoherence due to spontaneous decay from this level is
suppressed;

(ii) neither adjustment of the level spacings of qubit systems nor
adjustment of the cavity mode frequency during the gate operation is needed,
thus the operation is much simplified;

(iii) the operation time required for the gate realization is independent of
the number of qubits and thus does not increase with the number of qubits;

(iv) the phase shift on each target qubit can be adjusted by varying the
Rabi frequencies of the microwave pulses applied to the target qubit
systems; and

(v) the $n$ two-qubit CP gates forming this multiqubit phase gate can be
simultaneously performed using six operational steps (independent of $n$).
Thus, the gate operation is much simplified, especially when the number $n$
is large. In addition, as shown below, the present method can be applied for
implementing certain types of significant multiqubit phase gates, e.g., the
multiqubit phase gate consisting of $n$ two-qubit successive CP gates, which
are key elements in the quantum Fourier transform (QFT).

We stress that this proposal is quite general, and can be applied to
physical systems such as trapped atoms in a cavity and various types of
superconducting qubit systems coupled to a resonator. We note that
implementing a \textit{two-qubit} tunable phase gate with two atoms based on
cavity QED was previously proposed [33,34]. However, to the best of our
knowledge, no one has yet demonstrated how to perform a \textit{multiqubit}
tunable phase gate with one qubit simultaneously controlling $n$ qubits in
cavity QED. Also, our work is the first to show how to perform the $n$
successive two-qubit CP gates in QFT by the use of only \textit{one}
single-mode cavity. We believe that this proposal is useful since it
provides a simple and general protocol for realizing a multiqubit phase gate
for which the phase shift on each target qubit is \textit{tunable}.

This paper is organized as follows. In Sec. II, we briefly review the basic
theory of four-level quantum systems coupled to a single-mode cavity and/or
driven by classical pulses. In Sec. III, we show how to realize a multiqubit
tunable phase gate with one qubit simultaneously controlling $n$ target
qubits, by the use of $\left( n+1\right) $ qubit systems in a cavity. In
Sec.~IV, we discuss how to apply the present method for implementing two
types of significant quantum phase gates with multiple target qubits or
multiple control qubits. In Sec.~V, we give a brief discussion of the
experimental feasibility for implementing a six-qubit controlled-phase gate,
which are key elements in producing a QFT. In Sec. VI, we compare our work
with previous work. A concluding summary is presented in Sec. VII.

\begin{center}
\textbf{II. SYSTEMS INTERACTING WITH A CAVITY AND PULSES}
\end{center}

The four-level quantum systems throughout this paper could be either natural
atoms or \textit{artificial} atoms (e.g., superconducting devices), which
have the four levels shown in Fig.~3. Note that the four-level structure in
Fig.~3(a) applies to superconducting charge- qubit systems [1], the one in
Fig.~3(b) applies to phase-qubit systems [2,35], and the one in Fig.~3(c)
applies to flux-qubit systems [1,36]. In addition, the four-level structure
in Fig.~3(b) is also available in atoms. By adjusting the level spacings of
the qubit systems or the cavity mode frequency, the level $\left|
0\right\rangle $ can be made not to be affected by the cavity mode or the
classical pulses (see the discussion given in the appendix). Under this
condition, the level $\left| 0\right\rangle $ can be dropped from the
discussion below (also, see Fig. 4). We note that four levels are needed.
This is because (as explained in Sec. III) a three-level quantum system is
not enough to implement our desired gate.

\begin{figure}[tbp]
\includegraphics[bb=77 268 466 533, width=10.6 cm, clip]{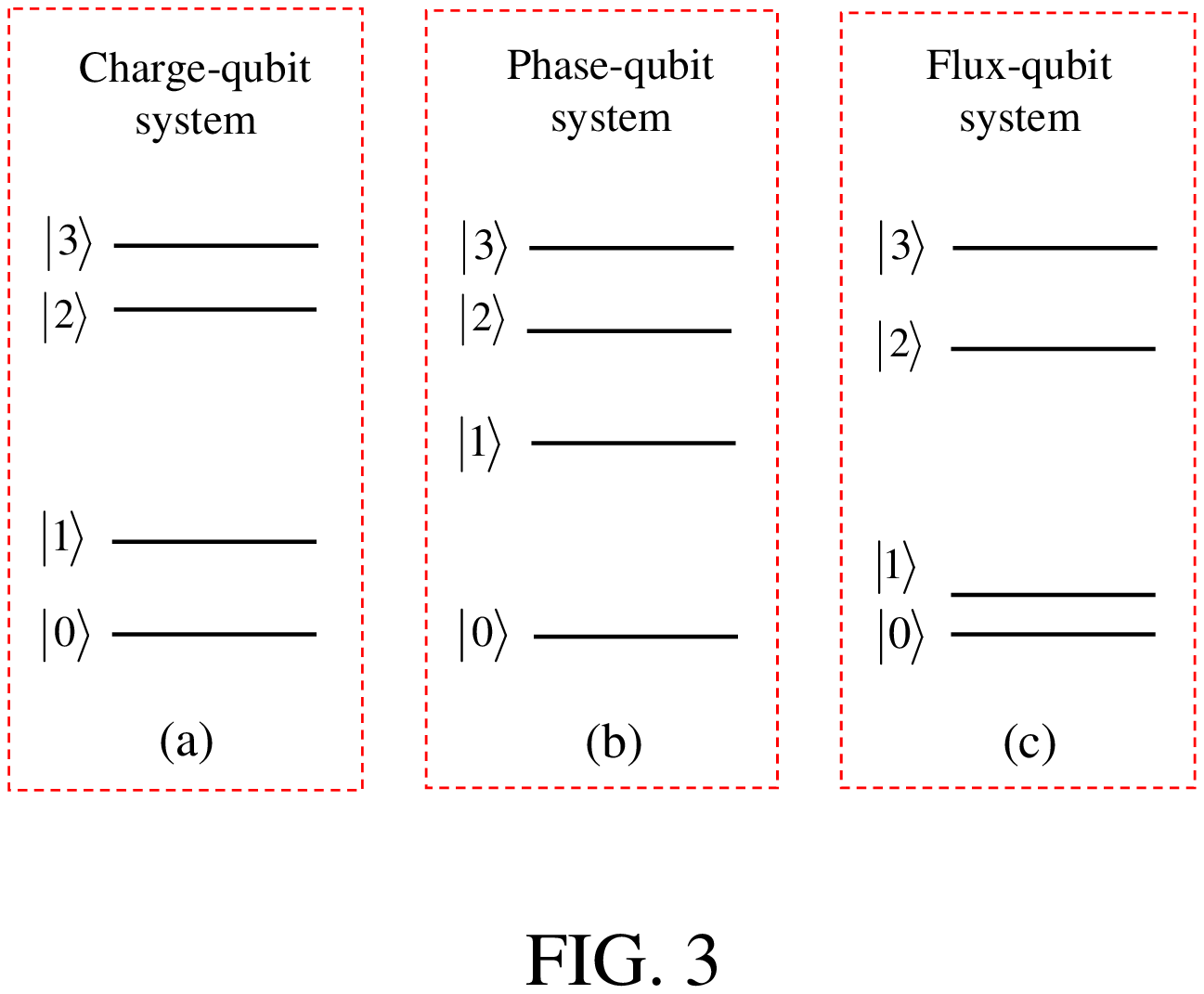} %
\vspace*{-0.08in}
\caption{Illustration of four-level qubit systems. The energy eigenvalues
for the four levels $\left| 0\right\rangle$, $\left| 1\right\rangle$, $%
\left| 2\right\rangle$, and $\left| 3\right\rangle$ are denoted by $E_0$, $%
E_1$, $E_2$, and $E_3$, respectively. In (a), the level spacings satisfy $%
E_2-E_1>E_1-E_0,~E_3-E_2$; and $E_3-E_2<E_1-E_0$. In (b), the level spacings
satisfy $E_1-E_0>E_2-E_1>E_3-E_2$. In (c), the level spacings meet $%
E_2-E_1>E_1-E_0,~E_3-E_2$; and $E_3-E_2>E_1-E_0$. The four-level structure
in (a) applies to charge-qubit systems, the one in (b) applies to
phase-qubit systems, and the one in (c) applies to flux-qubit systems. In
addition, the four-level structure in (b) is also available in atoms.}
\label{fig:3}
\end{figure}

\begin{center}
\textbf{A. System-cavity-pulse resonant Raman coupling}
\end{center}

Let us consider a qubit system in a cavity or coupled to a resonator, which
has a four-level structure depicted in Fig. 3. The energy eigenvalues for
the three levels $\left| 1\right\rangle $, $\left| 2\right\rangle ,$ and $%
\left| 3\right\rangle $ are denoted as $E_1,$ $E_2,$ and $E_3,$
respectively. Assume that the cavity mode is coupled to the $\left|
2\right\rangle \leftrightarrow \left| 3\right\rangle $ transition but
decoupled (highly detuned) from the transition between any other two levels
[Fig.~4(a)], which can be achieved by: (i) changing the cavity mode
frequency, (ii) choosing the qubit system (e.g., atom) appropriately to have
a desired level structure, and/or (iii) adjusting the level spacings of the
qubit system. Note that the cavity mode frequency for both optical cavities
and microwave cavities can be changed in various experiments (e.g., see,
[37-41]). And, for superconducting qubit systems, the level spacings can be
readily adjusted by varying the external parameters (e.g., the external
magnetic flux and gate voltage for superconducting charge-qubit systems, the
current bias or flux bias in the case of superconducting phase-qubit systems
and flux-qubit systems, see e.g. [1,2]). In addition, we assume that a
classical pulse is applied to the qubit system, which is coupled to the $%
\left| 1\right\rangle \leftrightarrow \left| 3\right\rangle $ transition but
decoupled (highly detuned) from the transition between any other two levels
[Fig.~4(a)]. The Hamiltonian of the whole system can thus be written as
\begin{eqnarray}
H &=&\hbar \omega _ca^{+}a+\sum_{l=1}^3E_l\left| l\right\rangle \left\langle
l\right| +\hbar g(a^{+}\sigma _{23}^{-}+\text{H.c.})  \nonumber \\
&&\ +\hbar \Omega (e^{i\omega t}\sigma _{13}^{-}+\text{H.c.}),
\end{eqnarray}
where $a^{+}$ and $a$ are the photon creation and annihilation operators of
the cavity mode with frequency $\omega _c$; $g$ is the coupling constant
between the cavity mode and the $\left| 2\right\rangle \leftrightarrow
\left| 3\right\rangle $ transition; $\Omega $ is the Rabi frequency of the
pulse, and $\omega $ is the frequency of the pulse; $\sigma _{23}^{-}=\left|
2\right\rangle \left\langle 3\right| ,$ and $\sigma _{13}^{-}=\left|
1\right\rangle \left\langle 3\right| .$

\begin{figure}[tbp]
\includegraphics[bb=71 218 396 633, width=8.6 cm, clip]{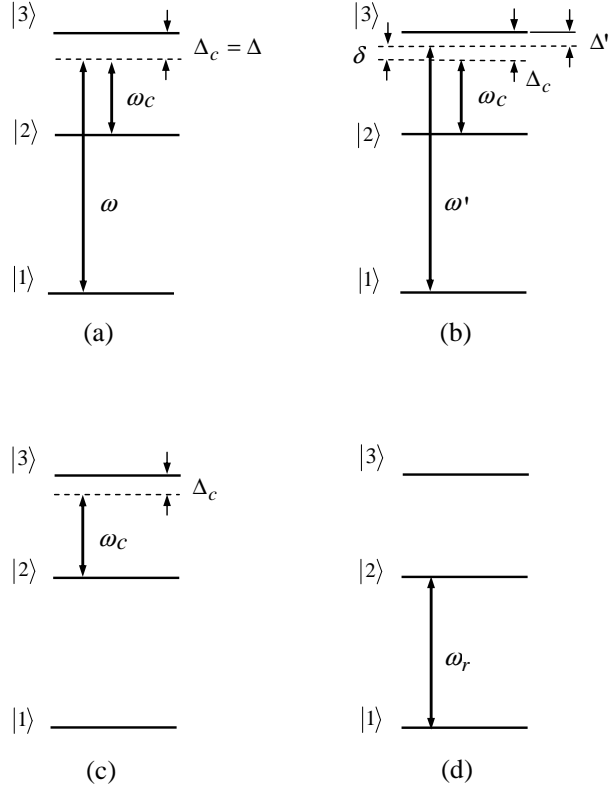} %
\vspace*{-0.08in}
\caption{(a) System-cavity-pulse resonant Raman coupling for a qubit system
in a cavity. Here, $\Delta=\omega _{31}-\omega$ is the detuning between the
pulse frequency $\omega$ and the $\left| 1\right\rangle \leftrightarrow
\left| 3\right\rangle $ transition frequency $\omega _{31}$, while $\Delta
_c=\omega _{32}-\omega _c$ is the detuning between the cavity mode frequency
$\omega _c$ and the $\left| 2\right\rangle \leftrightarrow \left|
3\right\rangle $ transition frequency $\omega _{32}$. The detunings $\Delta
_c$ and $\Delta$ are set to be equal, i.e., $\Delta _c=\Delta,$ in order to
establish the resonant Raman coupling between the two levels $\left|
1\right\rangle $ and $\left| 2\right\rangle $ of the qubit system. (b)
System-cavity-pulse off-resonant Raman coupling for $n$ qubit systems ($%
2,3,...,n+1$) in a cavity. For simplicity, we here only draw a figure for
qubit system $k$ interacting with the cavity mode and a classical microwave
pulse ($k=2,3,...,n+1$). Here, $\delta =\Delta _c-\Delta^{\prime }$ is the
detuning of the cavity mode with the pulse, $\Delta^{\prime }=\omega
_{31}-\omega^{\prime }$ is the detuning between the pulse frequency $%
\omega^{\prime }$ and the $\left| 1\right\rangle \leftrightarrow \left|
3\right\rangle $ transition frequency of qubit system $k.$ Note that $\delta
,$ $\Delta _c,$ and $\Delta^{\prime }$ are the same for all qubit systems ($%
2,3,...,n+1$) and that the pulses applied to the individual qubit systems ($%
2,3,...,n+1$) have the same frequency $\omega^{\prime }.$ The Rabi frequency
of the pulse applied to qubit system $k$ is denoted by $\Omega _k$. (c)
System-cavity off-resonant interaction for $n$ qubit systems ($2,3,...,n+1$)
in a cavity. The cavity mode is off-resonant with the $\left| 2\right\rangle
\leftrightarrow \left| 3\right\rangle $ transition of each qubit system,
with a detuning $\Delta _c$. (d) System-pulse resonant interaction. The
pulse with a frequency $\omega _r $ is resonant with the $\left|
1\right\rangle \leftrightarrow \left| 2\right\rangle $ transition.}
\label{fig:4}
\end{figure}

Let us now assume that the cavity mode is off-resonance with the $\left|
2\right\rangle \leftrightarrow \left| 3\right\rangle $ transition, i.e., $%
\Delta _c=\omega _{32}-\omega _c\gg g,$ and the pulse is off-resonance with
the $\left| 1\right\rangle \leftrightarrow \left| 3\right\rangle $
transition, i.e., $\Delta =\omega _{31}-\omega \gg \Omega $, where $\omega
_{32}=(E_3-E_2)/\hbar $ and $\omega _{31}=(E_3-E_1)/\hbar $ are respectively
the $\left| 2\right\rangle \leftrightarrow \left| 3\right\rangle $
transition frequency and the $\left| 1\right\rangle \leftrightarrow \left|
3\right\rangle $ transition frequency [Fig.~4(a)]. Under these conditions,
the level $\left| 3\right\rangle $ can be adiabatically eliminated [42].
Thus, for $\Delta _c=\Delta _{\mu w},$ the effective Hamiltonian in the
interaction picture is [43]
\begin{equation}
H_I=-\hbar \left[ \frac{\Omega ^2}\Delta \left| 1\right\rangle \left\langle
1\right| +\frac{g^2}{\Delta _c}a^{+}a\left| 2\right\rangle \left\langle
2\right| +\frac{\Omega g}{\Delta _c}(a^{+}\sigma _{12}^{+}+\text{H.c.}%
)\right]
\end{equation}
where $\sigma _{12}^{+}=\left| 2\right\rangle \left\langle 1\right| $. The
first two terms in Eq.~(4) are ac-Stark shifts of the levels $\left|
1\right\rangle $ and $\left| 2\right\rangle $ induced by the pulse and the
cavity mode, respectively; while the last two terms in Eq.~(4) are the
familiar Jaynes-Cummings interaction, describing the resonant Raman coupling
between the two levels $\left| 1\right\rangle $ and $\left| 2\right\rangle $%
, which results from the cooperation of the cavity mode and the pulse.

For the case of $\Omega =g,$ the initial states $\left| 2\right\rangle
\left| 1\right\rangle _c$ and $\left| 1\right\rangle \left| 0\right\rangle
_c $ of the whole system, under the Hamiltonian (4), evolve as follows
\begin{eqnarray}
\left| 2\right\rangle \left| 1\right\rangle _c &\rightarrow &e^{ig^2t/\Delta
_c}\left[ \cos \left( g^2t/\Delta _c\right) \left| 2\right\rangle \left|
1\right\rangle _c-i\sin \left( g^2t/\Delta _c\right) \left| 1\right\rangle
\left| 0\right\rangle _c\right] ,  \nonumber \\
\left| 1\right\rangle \left| 0\right\rangle _c &\rightarrow &e^{ig^2t/\Delta
_c}\left[ -i\sin \left( g^2t/\Delta _c\right) \left| 2\right\rangle \left|
1\right\rangle _c+\cos \left( g^2t/\Delta _c\right) \left| 1\right\rangle
\left| 0\right\rangle _c\right] ,
\end{eqnarray}
where $\left| 0\right\rangle _c$ and $\left| 1\right\rangle _c$ are the
vacuum state and the single-photon state of the cavity mode, respectively.
The state $\left| 0\right\rangle \left| 0\right\rangle _c$ remains unchanged
under the Hamiltonian (4).

\begin{center}
\textbf{B. System-cavity-pulse off-resonant Raman coupling}
\end{center}

Consider $n$ qubit systems labeled by $2,3,...,$ and $n+1$. Each qubit
system has a four-level structure, as described above. The cavity mode is
coupled to the $\left| 2\right\rangle \leftrightarrow \left| 3\right\rangle $
transition of each qubit system, but decoupled (highly detuned) from the
transition between any other two levels [Fig.~4(b)]. In addition, a
classical pulse is applied to each one of qubit systems ($2,3,...,n+1$),
which is coupled to the $\left| 1\right\rangle \leftrightarrow \left|
3\right\rangle $ transition but decoupled from the transition between any
other two levels [Fig.~4(b)]. Each pulse has the same frequency $\omega
^{\prime }$. The Hamiltonian for the whole system in the Schr\"odinger
picture is
\begin{equation}
H=\hbar \omega _ca^{+}a+\sum_{k=2}^{n+1}\left[ \sum_{l=1}^3E_l\left|
l\right\rangle _k\left\langle l\right| +\hbar g(a^{+}\sigma _{23,k}^{-}+%
\text{H.c.})+\hbar \Omega _k(e^{i\omega ^{\prime }t}\sigma _{13.k}^{-}+\text{%
H.c.})\right] ,
\end{equation}
where the subscript $k$ represents the $k$th qubit system$,$ $\Omega _k$ is
the Rabi frequency of the pulse applied to the $k$th qubit system, $\sigma
_{23,k}^{-}=\left| 2\right\rangle _k\left\langle 3\right| ,$ $\sigma
_{13,k}^{-}=\left| 1\right\rangle _k\left\langle 3\right| ,$ and $g$ is the
coupling constant between the cavity mode and the $\left| 2\right\rangle
\leftrightarrow \left| 3\right\rangle $ transition of each qubit system.

The detuning between the $\left| 1\right\rangle \leftrightarrow \left|
3\right\rangle $ transition frequency of the $k$th qubit system and the
frequency of the pulse applied to the $k$th qubit system is $\Delta ^{\prime
}=\omega _{31}-\omega ^{\prime }$ ($k=2,3,...,n+1$), which is identical for
each qubit system (i.e., independent of $k$) because of the same frequency $%
\omega ^{\prime }$ of each pulse [Fig.~4(b)]. Under the condition $\Delta
_c\gg g$ and $\Delta ^{\prime }\gg \max $ \{$\Omega _2,\Omega _3,...,\Omega
_{n+1}$\}$,$ the effective Hamiltonian in the interaction picture can be
written as [43]
\begin{eqnarray}
H_{\mathrm{eff}} &=&-\hbar \sum_{k=2}^{n+1}\left[ \frac{\Omega _k^2}{\Delta
^{\prime }}\left| 1\right\rangle _k\left\langle 1\right| +\frac{g^2}{\Delta
_c}a^{+}a\left| 2\right\rangle _k\left\langle 2\right| \right.  \nonumber \\
&&\ \left. +\chi _k(e^{-i\delta t}a^{+}\sigma _{12,k}^{+}+\text{H.c.}%
)\right] ,
\end{eqnarray}
where $\sigma _{12,k}^{+}=\left| 2\right\rangle _k\left\langle 1\right| ,$ $%
\delta =\Delta _c-\Delta ^{\prime },$ and
\begin{equation}
\chi _k=\Omega _kg(1/\Delta _c+1/\Delta ^{\prime })/2.
\end{equation}
For $\delta \gg g^2/\Delta _c,$ $\delta \gg \max $ \{$\Omega _2^2/\Delta
^{\prime },\Omega _3^2/\Delta ^{\prime },...,\Omega _{n+1}^2/\Delta ^{\prime
}$\}$,$ and $\delta \gg \max $\{$\chi _2,\chi _3,...,\chi _{n+1}\},$ there
is no energy exchange between the qubit systems and the cavity mode. Thus,
the effective Hamiltonian can be further written as [43-45]
\begin{eqnarray}
H_{\mathrm{eff}} &=&-\hbar \sum_{k=2}^{n+1}\left( \frac{\Omega _k^2}{\Delta
^{\prime }}\left| 1\right\rangle _k\left\langle 1\right| +\frac{g^2}{\Delta
_c}a^{+}a\left| 2\right\rangle _k\left\langle 2\right| \right)  \nonumber \\
&&\ -\hbar \sum_{k=2}^{n+1}\left[ \frac{\chi _k^2}\delta \left( a^{+}a\left|
2\right\rangle _k\left\langle 2\right| -aa^{+}\left| 1\right\rangle
_k\left\langle 1\right| \right) \right]  \nonumber \\
&&\ +\hbar \sum_{k\neq k^{\prime }=2}^{n+1}\frac{\chi _k\chi _{k^{\prime }}}%
\delta \left( \sigma _{12,k}^{+}\sigma _{12,k^{\prime }}^{-}+\sigma
_{12,k}^{-}\sigma _{12,k^{\prime }}^{+}\right) ,  \nonumber \\
&&
\end{eqnarray}
where the two terms in the second line above describe the photon-number
dependent Stark shifts induced by the off-resonant Raman coupling, and the
two terms in the last parentheses describe the ``dipole'' coupling between
the two qubit systems ($k,k^{\prime })$ mediated by the cavity mode and the
classical pulses. In the case when the level $\left| 1\right\rangle $ of
each qubit system is not populated, the Hamiltonian (9) reduces to
\begin{equation}
H_{\mathrm{eff}}=-\hbar \sum_{k=2}^{n+1}\left( \frac{g^2}{\Delta _c}+\frac{%
\chi _k^2}\delta \right) a^{+}a\left| 2\right\rangle _k\left\langle 2\right|
.
\end{equation}
The time-evolution operator for the Hamiltonian (10) is given by
\begin{equation}
U(t)=\otimes _{k=2}^{n+1}U_{kc}\left( t\right) ,
\end{equation}
where $U_{kc}\left( t\right) =\exp [i\left( g^2/\Delta _c+\chi _k^2/\delta
\right) a^{+}a\left| 2\right\rangle _k\left\langle 2\right| t]$ is the
time-evolution operator acting on the cavity mode and the $k$th qubit system$%
,$ which takes the following form
\begin{equation}
U_{kc}\left( t\right) =\left(
\begin{array}{cccc}
1 & 0 & 0 & 0 \\
0 & 1 & 0 & 0 \\
0 & 0 & 1 & 0 \\
0 & 0 & 0 & e^{i\varphi _k(t)}
\end{array}
\right)
\end{equation}
in the basis states $\left| 0\right\rangle _k\left| 0\right\rangle _c=\left(
1,0,0,0\right) ^T,$ $\left| 0\right\rangle _k\left| 1\right\rangle _c=\left(
0,1,0,0\right) ^T,$ $\left| 2\right\rangle _k\left| 0\right\rangle _c=\left(
0,0,1,0\right) ^T,$ and $\left| 2\right\rangle _k\left| 1\right\rangle
_c=\left( 0,0,0,1\right) ^T.$ Here, $\varphi _k\left( t\right) =\exp \left[
i\left( g^2/\Delta _c+\chi _k^2/\delta \right) t\right] .$ Note that the
conditional phase $\varphi _k$ is adjustable by varying the Rabi frequency $%
\Omega _k$ [see Eq. (8)].

\begin{center}
\textbf{C. System-cavity off-resonant interaction}
\end{center}

Consider $n$ qubit systems ($2,3,...,n+1$) with the four-level structure
described above. The cavity mode is coupled to the $\left| 2\right\rangle
\leftrightarrow \left| 3\right\rangle $ transition of each qubit system but
decoupled (highly detuned) from the transition between any other two levels
[Fig.~4(c)]. The Hamiltonian for the whole system is given by
\begin{equation}
H=\hbar \omega _ca^{+}a+\sum_{k=2}^{n+1}\sum_{l=1}^3E_l\left| l\right\rangle
_k\left\langle l\right| +\hbar \sum_{k=2}^{n+1}g(a^{+}\sigma _{23,k}^{-}+%
\text{H.c.}),
\end{equation}
where the subscript $k$ represents the $k$th qubit system$,$ $\sigma
_{23,k}^{-}=\left| 2\right\rangle _k\left\langle 3\right| $, and $l=0,1,2,3.$

In the interaction picture, the Hamiltonian (13) becomes

\begin{equation}
H=\hbar \sum_{k=2}^{n+1}g(e^{-i\Delta _ct}a^{+}\sigma _{23,k}^{-}+\text{H.c.}%
),
\end{equation}
For the case of $\Delta _c\gg g$ (i.e., the cavity mode is off-resonant with
the $\left| 2\right\rangle \leftrightarrow \left| 3\right\rangle $
transition of each qubit system), no energy exchange occurs between the
qubit systems and the cavity mode. Thus, the effective Hamiltonian can be
written as [44,45]
\begin{eqnarray}
H_{\mathrm{eff}} &=&-\hbar \sum_{k=2}^{n+1}\frac{g^2}{\Delta _c}\left(
a^{+}a\sigma _{22,k}-aa^{+}\sigma _{33,k}\right)  \nonumber \\
&&+\hbar \sum_{k\neq k^{\prime }=2}^{n+1}\frac{g^2}{\Delta _c}\left( \sigma
_{23,k}^{+}\sigma _{23,k^{\prime }}^{-}+\sigma _{23,k}^{-}\sigma
_{23,k^{\prime }}^{+}\right) ,  \nonumber \\
&&
\end{eqnarray}
where the two terms in the first line above represent the
photon-number-dependent Stark shifts, while the two terms in the second line
above describe the ``dipole'' coupling between the two qubit systems ($%
k,k^{\prime }$) mediated by the cavity mode. When the level $\left|
3\right\rangle $ of each qubit system is not excited, the Hamiltonian (15)
reduces to
\begin{equation}
H_{\mathrm{eff}}=-\hbar \sum_{k=2}^{n+1}\frac{g^2}{\Delta _c}a^{+}a\sigma
_{22,k}.
\end{equation}

The time-evolution operator for the Hamiltonian (16) is given by
\begin{equation}
\widetilde{U}(t)=\otimes _{k=2}^{n+1}\widetilde{U}_{kc}\left( t\right) ,
\end{equation}
where $\widetilde{U}_{kc}\left( t\right) =\exp [i\left( g^2/\Delta _c\right)
a^{+}a\sigma _{22,k}t]$ is the time-evolution operator acting on the cavity
mode and the $k$th qubit system. The operator $\widetilde{U}_{kc}\left(
t\right) $ is expressed as
\begin{equation}
\widetilde{U}_{kc}\left( t\right) =\left(
\begin{array}{cccc}
1 & 0 & 0 & 0 \\
0 & 1 & 0 & 0 \\
0 & 0 & 1 & 0 \\
0 & 0 & 0 & e^{i\phi _k(t)}
\end{array}
\right)
\end{equation}
in the basis states $\left| 0\right\rangle _k\left| 0\right\rangle _c=\left(
1,0,0,0\right) ^T,$ $\left| 0\right\rangle _k\left| 1\right\rangle _c=\left(
0,1,0,0\right) ^T,$ $\left| 2\right\rangle _k\left| 0\right\rangle _c=\left(
0,0,1,0\right) ^T,$ and $\left| 2\right\rangle _k\left| 1\right\rangle
_c=\left( 0,0,0,1\right) ^T.$ Here, $\phi _k\left( t\right) =\exp \left(
ig^2t/\Delta _c\right) .$

\begin{center}
\textbf{D. System-pulse resonant interaction}
\end{center}

Consider now a qubit system driven by a classical pulse with frequency $%
\omega _r$ and initial phase $\phi $. Moreover, assume that the pulse is
resonant with the $\left| 1\right\rangle \leftrightarrow \left|
2\right\rangle $ transition, i.e., $\omega _r=\omega _{21},$ where $\omega
_{21}=\left( E_2-E_1\right) /\hbar $ is the transition frequency between the
two levels $\left| 1\right\rangle $ and $\left| 2\right\rangle $
[Fig.~4(d)]. In this case, the interaction Hamiltonian in the interaction
picture is given by
\begin{equation}
H_I=\frac \hbar 2\left( \widetilde{\Omega }e^{i\phi }\left| 1\right\rangle
\left\langle 2\right| +\text{H.c.}\right) ,
\end{equation}
where $\widetilde{\Omega }$ is the Rabi frequency of the pulse. From the
Hamiltonian (19), it is straightforward to show that a pulse of duration $t$
results in the following rotation
\begin{eqnarray}
\left| 1\right\rangle &\rightarrow &\cos \frac{\widetilde{\Omega }}2t\left|
1\right\rangle -ie^{-i\phi }\sin \frac{\widetilde{\Omega }}2t\left|
2\right\rangle ,  \nonumber \\
\left| 2\right\rangle &\rightarrow &-ie^{i\phi }\sin \frac{\widetilde{\Omega
}}2t\left| 1\right\rangle +\cos \frac{\widetilde{\Omega }}2t\left|
2\right\rangle .
\end{eqnarray}

Above we have introduced four types of interaction of qubit systems with the
cavity mode and/or the pulses. The results presented above will be employed
for the gate implementation discussed in next section.

\begin{center}
\textbf{III. IMPLEMENTATION OF MULTIQUBIT PHASE GATES}
\end{center}

Let us now consider $(n+1)$ identical qubit systems in a single-mode cavity
or coupled to a resonator, which are labelled by 1,~2,~...,~and~$n+1,$
respectively. The qubit systems ($1,2,...,n+1$) each have the four-level
configuration depicted in Fig.~3. For each qubit system, the two lowest
levels $\left| 0\right\rangle $ and $\left| 1\right\rangle $ represent the
two logical states of a qubit while the two higher-energy intermediate
levels $\left| 2\right\rangle $ and $\left| 3\right\rangle $ are used for
the gate implementation. In the following, qubit system $1$ acts as a
\textit{control} while each one of the qubit systems ($2,3,...,n+1$) plays a
\textit{target} role.

\begin{figure}[tbp]
\includegraphics[bb=60 83 510 781, width=10.4 cm, clip]{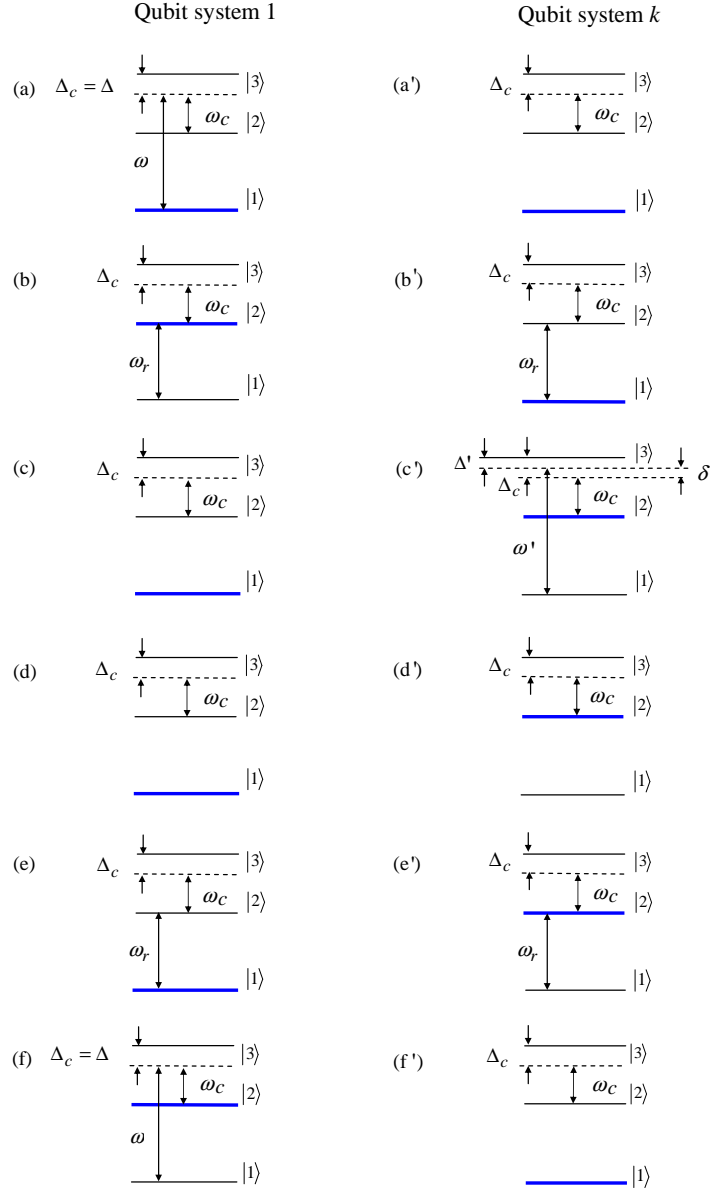} %
\vspace*{-0.08in}
\caption{(color online) Illustration of qubit systems interacting with the
cavity mode and/or the microwave pulses for each step of operations during
the gate performance. In each figure, blue lines represent the level
population of qubit systems \textit{before each step of operations}. Figures
from top to bottom correspond to the operations of steps (i)$\sim$(vi),
respectively. The figures on the left side correspond to qubit system $1$,
while the figures on the right side correspond to qubit system $k,$ with $%
k=2,3,...,n+1$. In each figure, $\Delta _c=\omega _{32}-\omega _c$ is the
detuning between the cavity mode frequency $\omega _c$ and the $\left|
2\right\rangle \leftrightarrow \left| 3\right\rangle $ transition frequency $%
\omega _{32}$. In (a) and (f), $\Delta=\omega _{31}-\omega$ is the detuning
between the $\left| 1\right\rangle \leftrightarrow \left| 3\right\rangle $
transition frequency $\omega _{31}$ and the pulse frequency $\omega$. In (c$%
^{\prime }$), $\Delta^{\prime }=\omega _{31}-\omega^{\prime }$ is the
detuning between the $\left| 1\right\rangle \leftrightarrow \left|
3\right\rangle $ transition frequency $\omega _{31}$ and the frequency $%
\omega^{\prime }$ of the pulse applied to qubit system $k,$ and $\delta
=\Delta _c-\Delta^{\prime }$ is the detuning of the cavity mode with the
pulse.}
\label{fig:5}
\end{figure}

To realize the proposed gate, let us assume that the level $\left|
0\right\rangle $ of each qubit system is not affected by the cavity mode
and/or the pulses during the gate operation; and the cavity mode is
off-resonance with the $\left| 2\right\rangle \leftrightarrow \left|
3\right\rangle $ transition of each qubit system with a detuning $\Delta _c$
but decoupled (highly detuned) from the transition between any other two
levels during the entire operation below. As mentioned above, these
conditions can be achieved by prior adjustment of the cavity mode frequency
[37-41] or the level spacings of the qubit systems before the gate operation
[1,2,46], or by appropriately choosing the qubit systems (i.e., atoms) to
have the desired level structure. In addition, assume that quantum
information is initially stored by the two lowest levels $\left|
0\right\rangle $ and $\left| 1\right\rangle $ of each qubit system and the
cavity mode is initially in the vacuum state $\left| 0\right\rangle _c.$

The procedure for realizing the ($n+1$)-qubit phase gate described by
Eq.~(2) is as follows:

\textit{Step (i)}: Apply a classical pulse to qubit system $1$ to induce the
Raman transition described in Sec. II A [Fig.~5(a)]. It can be seen from
Eq.~(5) that after a pulse duration $t_1=\pi \Delta _c/\left( 2g^2\right) ,$
the state $\left| 1\right\rangle _1\left| 0\right\rangle _c$ for qubit
system $1$ and the cavity mode is transformed to the state $\left|
2\right\rangle _1\left| 1\right\rangle _c.$ Namely, when the qubit system $1$
is initially in the state $\left| 1\right\rangle $, a photon is emitted to
the cavity mode after the pulse. On the other hand, the state $\left|
0\right\rangle _1\left| 0\right\rangle _c$ remains unchanged during the
pulse.

\textit{Step (ii)}: Apply a classical pulse (with $\omega _r=\omega _{21}$
and $\phi =\pi /2$) to qubit system $1$ [Fig.~5(b)] and a classical pulse
(with $\omega _r=\omega _{21}$ and $\phi =-\pi /2$) to each of qubit systems
($2,3,...,n+1$) [Fig.~5(b$^{\prime }$)]. The duration of each pulse is set
by $t_2=\pi /\left( 2\widetilde{\Omega }\right) $ (here, $\widetilde{\Omega }
$ is the Rabi frequency of each pulse). It can be seen from Eq.~(20) that
after the pulses, the state $\left| 2\right\rangle $ of qubit system $1$ is
transformed to the state $\left| 1\right\rangle $ while the state $\left|
1\right\rangle $ of each of qubit systems ($2,3,...,n+1$) is transformed to
the state $\left| 2\right\rangle $. Note that since both of the two levels $%
\left| 2\right\rangle $ and $\left| 3\right\rangle $ of qubit system $1$ are
unpopulated after the operation of this step, \textit{the qubit system $1$
is decoupled from the cavity mode during the operations of steps (iii) and
(iv) below} [see Figs. 5(c,d)].

\textit{Step (iii)}: Apply a classical pulse (with duration $t_3$) to each
of qubit systems ($2,3,...,n+1$) to induce the conditional phase shift
described in Sec. II B [Fig.~5(c$^{\prime }$)]. Since the level $\left|
1\right\rangle $ for qubit systems ($2,3,...,n+1$) is not populated after
step (ii) above, the effective Hamiltonian of the whole system is given by
Eq.~(10). Accordingly, the time-evolution operator describing this step is
the operator given by Eq.~(11) for $t=t_3,$ i.e.,
\begin{equation}
U(t_3)=\otimes _{k=2}^{n+1}U_{kc}\left( t_3\right) ,
\end{equation}
where $U_{kc}\left( t_3\right) $ is the time-evolution operator acting on
the cavity mode and qubit system $k$ ($k=2,3,...,n+1$), which takes the
matrix of Eq. (12) for $t=t_3$.

\textit{Step (iv)}: Wait for a time $t_{4.}$ Since no pulse is applied to
each qubit system and the cavity mode is off-resonance with the $\left|
2\right\rangle \leftrightarrow \left| 3\right\rangle $ transition of each of
qubit systems ($2,3,...,n+1$) [Fig.~5(d$^{\prime }$)], this is the case
discussed in Sec. II C. Note that the level $\left| 3\right\rangle $ of each
qubit system is not excited. Thus, the effective Hamiltonian of the whole
system is given by Eq.~(16). Consequently, the time-evolution operator
describing this step is the operator given by Eq.~(17) for $t=t_4,$ i.e.,
\begin{equation}
\widetilde{U}(t_4)=\otimes _{k=2}^{n+1}\widetilde{U}_{kc}\left( t_4\right) ,
\end{equation}
where $\widetilde{U}_{kc}\left( t_4\right) $ is the time-evolution operator
acting on the cavity mode and qubit system $k$, taking the matrix of Eq.
(18) for $t=t_4.$

Combining Eq.~(22) with Eq.~(21) leads to
\begin{eqnarray}
\left| 2\right\rangle _k\left| 1\right\rangle _c &\rightarrow &\exp \left[
ig^2\left( t_3+t_4\right) /\Delta _c\right]  \nonumber \\
&&\otimes \exp \left[ i\chi _k^2t_3/\delta \right] \left| 2\right\rangle
_k\left| 1\right\rangle _c.
\end{eqnarray}
When $t_3+t_4=2m\pi \Delta _c/g^2$ ($m$ is an integer), the transformation
(23) becomes
\begin{equation}
\left| 2\right\rangle _k\left| 1\right\rangle _c\rightarrow \exp \left(
i\theta _k\right) \left| 2\right\rangle _k\left| 1\right\rangle _c,\text{ }
\end{equation}
with
\begin{equation}
\theta _k=\frac{\chi _k^2}\delta t_3=\frac{\Omega _k^2g^2}{4\delta }\left(
\frac 1{\Delta _c}+\frac 1{\Delta ^{\prime }}\right) ^2t_3.
\end{equation}
Note that $g,$ $\delta ,$ $\Delta _c,$ and $\Delta ^{\prime }$ are the same
for \textit{all} qubit systems ($2,3,...,n+1$). Therefore, the angle $\theta
_k$ can be adjusted by changing the Rabi frequency $\Omega _k$ of the pulse
applied to the qubit system $k$. The result (24) shows that after the
operations of steps (i)$\sim$(iv), a phase shift $\exp \left( i\theta
_k\right) $ happens to the state $\left| 2\right\rangle $ of the qubit
system $k,$ if and only if the cavity mode is in the single-photon state $%
\left| 1\right\rangle _c.$

\textit{Step (v):} Apply a classical pulse (with $\omega _r=\omega _{21}$
and $\phi =-\pi /2$) to qubit system $1$ [Fig.~5(e)] and a classical pulse
(with $\omega _r=\omega _{21}$ and $\phi =\pi /2$) to each of qubit systems (%
$2,3,...,n+1$) [Fig.~5(e$^{\prime }$)]. After a pulse duration $t_2=\pi
/\left( 2\widetilde{\Omega }\right) ,$ the state $\left| 1\right\rangle $ of
qubit system $1$ is transformed to the state $\left| 2\right\rangle $ while
the state $\left| 2\right\rangle $ of qubit systems ($2,3,...,n+1$) is
transformed back to the original state $\left| 1\right\rangle .$

\textit{Step (vi):} Repeat the operation of step (i) [Fig.~5(f)]. After a
pulse duration $t_1=\pi \Delta _c/\left( 2g^2\right) $, the state $\left|
2\right\rangle _1\left| 1\right\rangle _c$ of the qubit system $1$ and the
cavity mode is changed to the state $\left| 1\right\rangle _1\left|
0\right\rangle _c,$ which shows that the state $\left| 2\right\rangle $ of
the qubit system $1$ is transformed back to the state $\left| 1\right\rangle
$ and the cavity mode returns to its original vacuum state. However, the
state $\left| 0\right\rangle _1\left| 0\right\rangle _c$ remains unchanged
during the pulse.

One can check that the ($n+1$)-qubit phase gate of one qubit simultaneously
controlling $n$ qubits, described by Eq.~(2), was obtained with ($n+1$)
qubit systems [i.e., the control qubit system 1 and the target qubit systems
($2,3,...,n+1$)] after the above manipulation.

To better see how the multiqubit phase gate described by Eq.~(2) is
implemented by the operations above, let us consider a three-qubit example.
One can check that the states of the whole system after each step of the
above operations are summarized below:

\[
\begin{array}{c}
\left| 100\right\rangle \left| 0\right\rangle _c \\
\left| 101\right\rangle \left| 0\right\rangle _c \\
\left| 110\right\rangle \left| 0\right\rangle _c \\
\left| 111\right\rangle \left| 0\right\rangle _c
\end{array}
\stackrel{\text{Step\thinspace (i)}}{\longrightarrow }
\begin{array}{c}
\left| 200\right\rangle \left| 1\right\rangle _c \\
\left| 201\right\rangle \left| 1\right\rangle _c \\
\left| 210\right\rangle \left| 1\right\rangle _c \\
\left| 211\right\rangle \left| 1\right\rangle _c
\end{array}
\stackrel{\text{Step\thinspace (ii)}}{\longrightarrow }
\begin{array}{c}
\left| 100\right\rangle \left| 1\right\rangle _c \\
\left| 102\right\rangle \left| 1\right\rangle _c \\
\left| 120\right\rangle \left| 1\right\rangle _c \\
\left| 122\right\rangle \left| 1\right\rangle _c
\end{array}
\]
\begin{eqnarray}
&&\stackrel{\text{Steps\thinspace (iii) and (iv)}}{\longrightarrow }
\begin{array}{c}
\;\;\;\;\;\;\;\;\;\,\;\;\;\left| 100\right\rangle \left| 1\right\rangle _c
\\
\;\;\;\;\;\;e^{i\theta _3}\left| 102\right\rangle \left| 1\right\rangle _c
\\
\;\;\;\;\;\;e^{i\theta _2}\left| 120\right\rangle \left| 1\right\rangle _c
\\
e^{i\theta _2}e^{i\theta _3}\left| 122\right\rangle \left| 1\right\rangle _c
\end{array}
\stackrel{\text{Step\thinspace (v)}}{\longrightarrow }
\begin{array}{c}
\;\;\;\;\;\;\;\;\,\;\;\;\;\left| 200\right\rangle \left| 1\right\rangle _c
\\
\;\;\;\;\;\;e^{i\theta _3}\left| 201\right\rangle \left| 1\right\rangle _c
\\
\;\;\;\;\;\;e^{i\theta _2}\left| 210\right\rangle \left| 1\right\rangle _c
\\
e^{i\theta _2}e^{i\theta _3}\left| 211\right\rangle \left| 1\right\rangle _c
\end{array}
\stackrel{\text{Step\thinspace (vi)}}{\longrightarrow }
\begin{array}{c}
\;\;\;\;\;\;\;\;\;\;\;\,\left| 100\right\rangle \left| 0\right\rangle _c \\
\;\;\;\;\;\;e^{i\theta _3}\left| 101\right\rangle \left| 0\right\rangle _c
\\
\;\;\;\;\;\;e^{i\theta _2}\left| 110\right\rangle \left| 0\right\rangle _c
\\
e^{i\theta _2}e^{i\theta _3}\left| 111\right\rangle \left| 0\right\rangle _c
\end{array}
,  \nonumber \\
&&
\end{eqnarray}
where $\left| ijk\right\rangle $ is an abbreviation of the state $\left|
i\right\rangle _1\left| j\right\rangle _2\left| k\right\rangle _3$ of qubit
systems ($1,2,3$) with $i,j,k\in \{0,1,2\}.$

On the other hand, it is obvious that the following states of the whole
system
\begin{equation}
\left| 000\right\rangle \left| 0\right\rangle _c,\;\left| 001\right\rangle
\left| 0\right\rangle _c,\;\left| 010\right\rangle \left| 0\right\rangle
_c,\;\left| 011\right\rangle \left| 0\right\rangle _c
\end{equation}
remain unchanged during the entire operation. Hence, it can be concluded
from Eq.~(26) that a three-qubit phase gate of one qubit simultaneously
controlling two qubits, described by Eq.~(2) (with $n=2$), was achieved with
three qubit systems (i.e., the control qubit system $1$ and the two target
qubit systems $2$ and $3$) after the above process.

During the operation of step (ii) or step (v), a single photon is populated
in the cavity mode and the state $\left| 2\right\rangle $ of each qubit
system is occupied. Therefore, for these two-step operations, there is an
accumulated phase shift $\exp (i2g^2t_2/\Delta _c)$ to the state $\left|
2\right\rangle $ of each qubit system. However, when $2t_2\ll t_3+t_4$, this
unwanted phase shift is sufficiently small and thus can be neglected. Note
that $t_2=\pi /\widetilde{\Omega }$ and $t_3+t_4=2m\pi \Delta _c/g^2.$
Therefore, the condition $2t_2\ll t_3+t_4$ turns into $\widetilde{\Omega }%
\gg $ $g^2/\left( m\Delta _c\right) $, which can be met by increasing the
pulse Rabi frequency $\widetilde{\Omega }$ (i.e., by increasing the
intensity of the resonant pulses).

Finally, we should mention that the method proposed here does not work for
qubit systems with three levels. Suppose now that the quantum information is
encoded in the states $\left| 1\right\rangle $ and $\left| 2\right\rangle .$
In this case, during the pulse applied to the control qubit system for
creating a cavity photon, one cannot have the target qubit systems decoupled
from the cavity mode.

\begin{center}
\textbf{IV. TWO TYPES OF MULTIQUBIT PHASE GATES}
\end{center}

Above we have shown how to realize the $(n+1)$-qubit phase gates described
by the operator (2). In this section, we focus on two types of multiqubit
phase gates, which are depicted in Fig.~6(a) and Fig.~7(a), respectively. It
can be seen from Fig.~6(a) and Fig.~7(a) that the two types of multiqubit
phase gates considered here consist of $n$ two-qubit controlled-phase gates
each having a \textit{shared} control qubit but a \textit{different} target
qubit. Below we give a discussion of their implementation and importance in
quantum information processing.

\begin{center}
\textbf{A. Implementation of gates of the first-type}
\end{center}

The controlled-phase gate with $n$-target qubits shown in Fig.~6(a) has the
property that the phase for the state $\left| 1\right\rangle $ of each
target qubit is shifted by the same amount $\pi ,$ when the control qubit
(i.e., qubit $1$) is in the state $\left| 1\right\rangle .$ Hence, this
multiqubit phase gate is a special one for $\theta _k=\pi $ above and thus
the procedure for realizing it is the same as that discussed in the previous
section. To obtain this multiqubit phase gate, one needs to set the Rabi
frequencies of the pulses applied to the target qubit systems ($2,3,...,n+1$%
) to be \textit{identical} (i.e., $\Omega _2=\Omega _3=...=\Omega _{n+1},$
leading to $\chi _2=\chi _3=...=\chi _{n+1}\equiv \chi $ ) and set the
interaction time $t_3$ above to be $t_3=\pi \delta /\chi ^2,$ such that $%
\theta _k=\chi _k^2t_3/\delta =\pi $ ($k=2,3,...,n+1$).

\begin{figure}[tbp]
\includegraphics[bb=60 284 544 605, width=10.6 cm, clip]{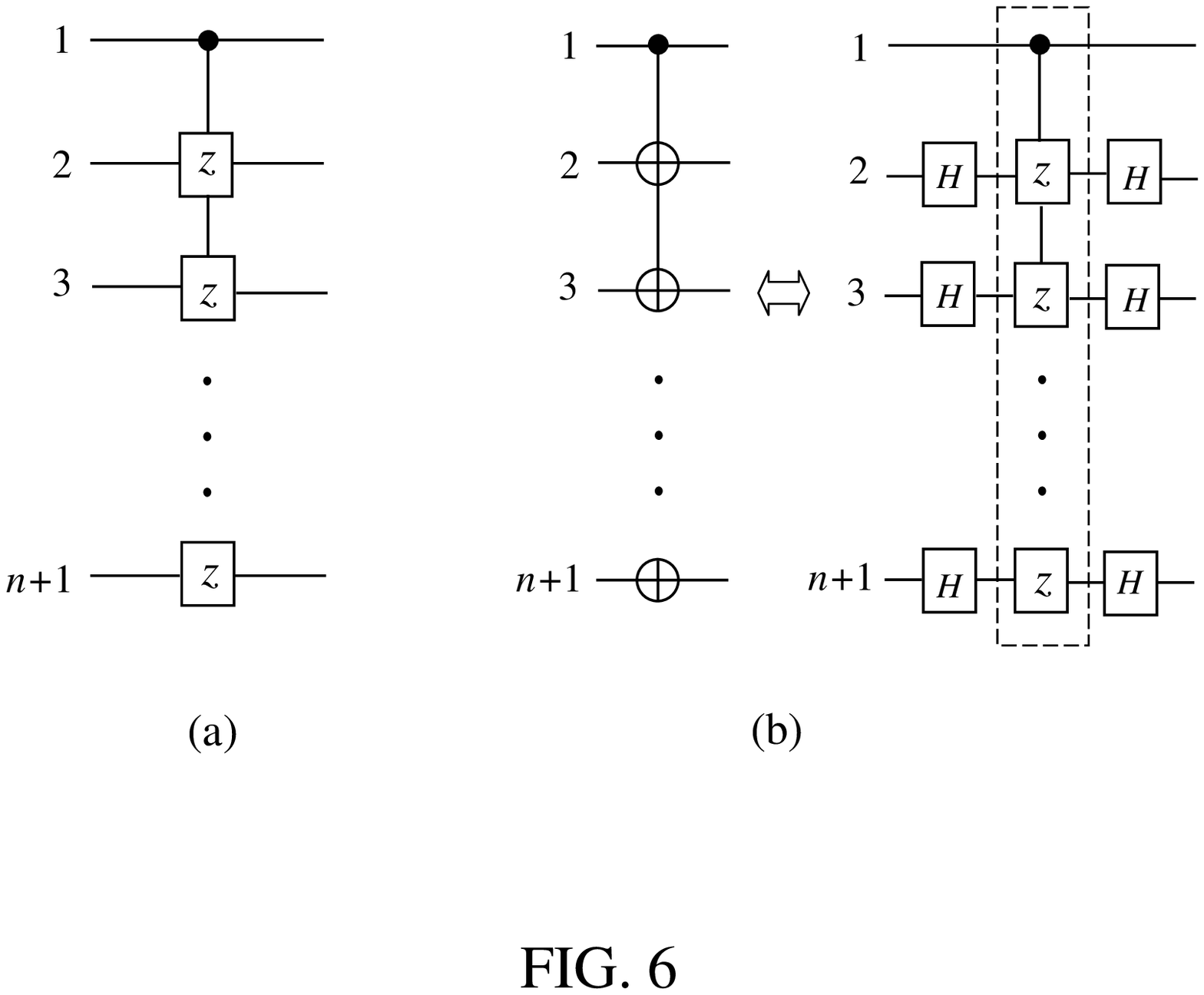} %
\vspace*{-0.08in}
\caption{(a) Circuit for the first type of controlled-phase gate with qubit $%
1$ simultaneously controlling $n$ target qubits ($2,~3,~...,$~$n+1$). Here, $%
Z$ represents a controlled-phase shift $e^{i\pi }$ on each target
qubit.~Namely, if the control qubit $1$ is in the state $\left|
1\right\rangle $, then the state $\left| 1\right\rangle $ at each $Z$ is
phase-shifted by $\pi$ (i.e., $\left| 1\right\rangle $ $\rightarrow $ $%
e^{i\pi}\left| 1\right\rangle=-\left| 1\right\rangle $), while the state $%
\left| 0\right\rangle $ remains unchanged.~(b) Relation between a $n$%
-target-qubit controlled-NOT gate and a $n$-target-qubit controlled-phase
gate. The circuit on the left side is equivalent to the circuit on the right
side. For the circuit on the left side, the symbol $\oplus $ represents a
CNOT gate on each target qubit.~If the control qubit $1$ is in the state $%
\left| 1\right\rangle $, then the state at each $\oplus $ is bit-flipped as $%
\left| 1\right\rangle $ $\rightarrow \left| 0\right\rangle $ and $\left|
0\right\rangle $ $\rightarrow \left| 1\right\rangle $.~However,~when the
control qubit $1$ is in the state $\left| 0\right\rangle $,~the state at
each $\oplus $ remains unchanged. On the other hand, for the circuit on the
right side, the part enclosed in the dashed-line box represents a $n$%
-target-qubit controlled-phase gate. The element containing $H$ corresponds
to a Hadamard transformation described by $\left| 0\right\rangle \rightarrow
\left( 1/\protect\sqrt{2}\right) \left( \left| 0\right\rangle +\left|
1\right\rangle \right) $ while $\left| 1\right\rangle \rightarrow \left( 1/%
\protect\sqrt{2}\right) \left( \left| 0\right\rangle -\left| 1\right\rangle
\right) .$}
\label{fig:6}
\end{figure}

\begin{figure}[tbp]
\includegraphics[bb=29 298 567 632, width=10.6 cm, clip]{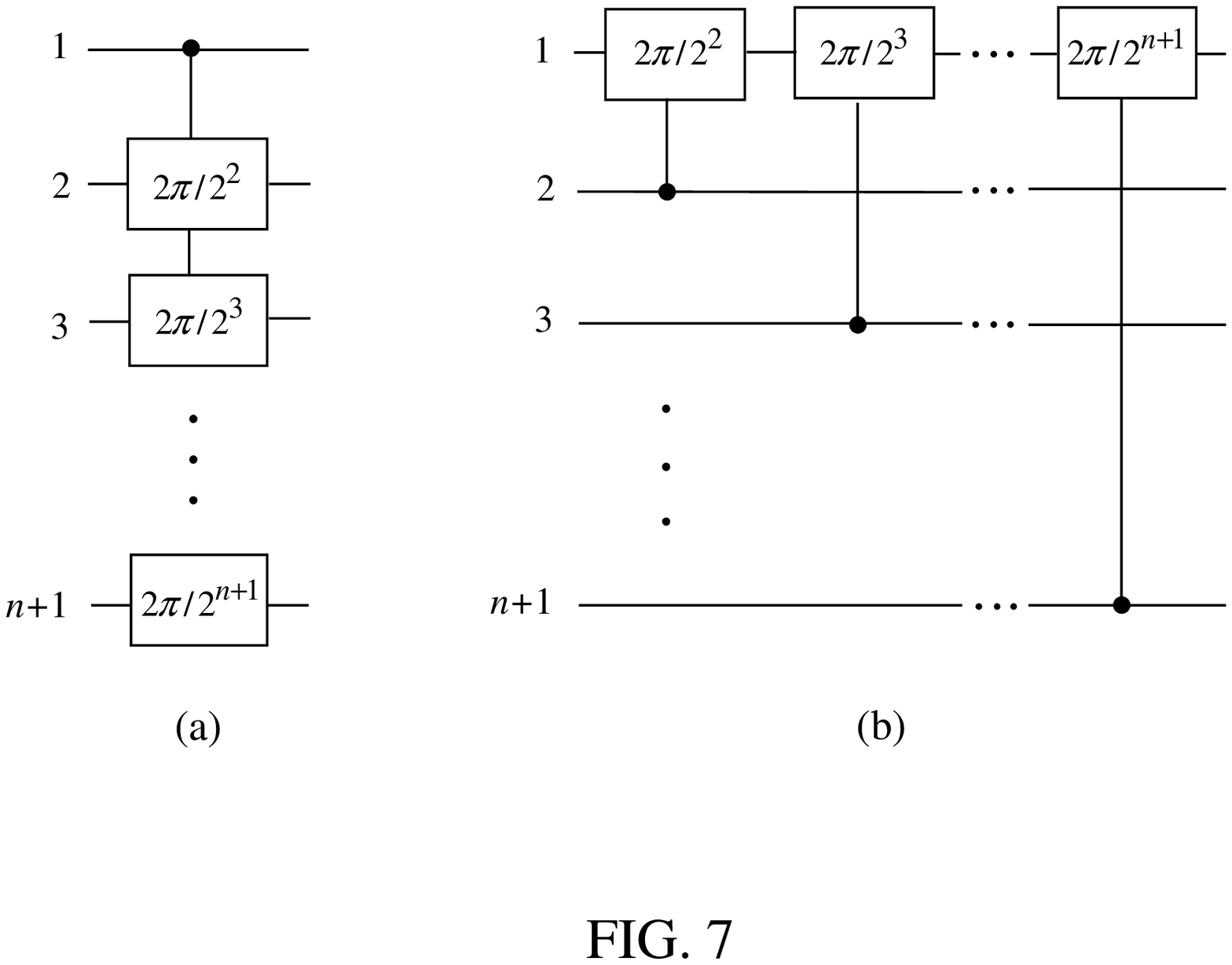} %
\vspace*{-0.08in}
\caption{(a) Circuit for the second type of controlled-phase gate with qubit
$1$ simultaneously controlling $n$ target qubits ($2,~3,~...,$~$n+1$). Here,
the element $2\pi /2^k$ represents a controlled-phase shift $\exp \left(
i2\pi /2^k\right)$, which is performed on the target qubit $k$ when the
control qubit $1$ is in the state $\left| 1\right\rangle $ ($k=2,3,...,n+1$%
). Namely, if the control qubit $1$ is in the state $\left| 1\right\rangle $%
, then a phase shift $2\pi /2^k$ occurs to the state $\left| 1\right\rangle $
of the target qubit $k$, while nothing happens otherwise. (b) Circuit for
the $n$ successive two-qubit CP gates in a quantum Fourier transform, which%
\textit{\ is equivalent to the circuit in (a).} Here, each two-qubit CP gate
has a shared target qubit (i.e., qubit $1$) but a different control qubit
(i.e., qubit $2,$ $3,...,$ or$,$ $n+1$). The element $2\pi /2^k$ represents
a controlled-phase shift $\exp \left( i2\pi /2^k\right)$, which is performed
on the target qubit $1$ when the control qubit $k$ is in the state $\left|
1\right\rangle $ ($k=2,3,...,n+1$). Namely, if the control qubit $k$ is in
the state $\left| 1\right\rangle $, then a phase shift by $2\pi /2^k$ occurs
to the state $\left| 1\right\rangle $ of the target qubit $1,$ while nothing
happens otherwise.}
\label{fig:7}
\end{figure}

Note that a controlled-NOT gate with $n$-target qubits, shown in Fig.~6(b),
can also be achieved using the present proposal. This is because this gate
is equivalent to the $n$-target-qubit controlled-phase gate discussed here,
plus a single-qubit Hadamard gate acting on each target qubit before and
after the $n$-target-qubit controlled-phase gate [Fig.~6(b)]

The controlled-phase (or controlled-NOT) gates with multiple target qubits
shown in Fig.~6(a) [or Fig.~6(b)] are important in quantum information
processing because they can be applied in entanglement preparation~[47],
error correction~[48], quantum algorithms (e.g., the Discrete Cosine
Transform~[49]), and quantum cloning~[50].

\begin{center}
\textbf{B. Implementation of gates of the second-type}
\end{center}

The controlled-phase gate with $n$-target qubits shown in Fig.~7(a) has the
property that the phase for the state $\left| 1\right\rangle $ of the target
qubit $k$ is shifted by $2\pi /2^k,$ when the control qubit (i.e., qubit $1$%
) is in the state $\left| 1\right\rangle .$ Thus, this multiqubit phase gate
is a special one for $\theta _k=$ $2\pi /2^k$ above and therefore can be
implemented using the procedure introduced in the previous section. To
realize it, the Rabi frequencies $\Omega _2,$ $\Omega _3,...,$ and $\Omega
_{n+1}$ for the pulses (respectively applied to the target qubit systems $%
2,3,...,$ and $n+1$) need to be set non-identical, and satisfy the relation $%
\Omega _{k+1}/\Omega _k=1/\sqrt{2},$ that is, $\chi _{k+1}/\chi _k=1/\sqrt{2}
$ ($k=2,3,...,n$); and the interaction time $t_3$ above needs to be set by $%
t_3=\left( \pi /2\right) \left( \delta /\chi _2^2\right) $. In this way, we
can obtain $\theta _k=\chi _k^2t_3/\delta =2\pi /2^k$ ($k=2,3,...,n+1$).

For any two-qubit CP gate described by the transformation $\left|
00\right\rangle \rightarrow \left| 00\right\rangle ,$ $\left|
01\right\rangle \rightarrow \left| 01\right\rangle ,$ $\left|
10\right\rangle \rightarrow \left| 10\right\rangle ,$ and $\left|
11\right\rangle \rightarrow e^{i\varphi }\left| 11\right\rangle ,$ it is
clear that the roles of the two qubits can be interchanged. Namely, the
first qubit can be either the control qubit or the target qubit, and the
same applies to the second qubit. One can see that when the second (first)
qubit is a control qubit, while the first (second) qubit is a target, the
phase of the state $\left| 1\right\rangle $ of the first (second) qubit is
shifted by $e^{i\varphi }$ when the second (first) qubit is in the state $%
\left| 1\right\rangle ,$ while nothing happens otherwise. Thus, the quantum
circuit in Fig.~7(a) is equivalent to the quantum circuit shown in
Fig.~7(b). It can be seen from Fig.~7(b) that each of the $n$ two-qubit CP
gates has a shared target qubit (i.e., qubit $1$) but a different control
qubit (i.e., qubit $2,$ $3,$ $...,$ or $n+1$). Note that the $n$ successive
two-qubit CP gates shown in Fig.~7(b) are key elements in QFT [51,52].
Hence, the importance for this second type of multiqubit controlled-phase
gate is obvious.

The above discussion shows that the two types of multiqubit controlled-phase
gates above can be implemented by appropriately setting the Rabi frequencies
of the pulses applied to the qubit systems. In addition, we should point out
that by adjusting the Rabi frequencies of the pulses (i.e., changing the
intensities of the pulses) applied to the target qubit systems, other types
of quantum controlled-phase gates with one qubit simultaneously controlling
multiple target qubits, which may have applications in quantum information
processing, can in principle be performed using this proposal.

\begin{center}
\textbf{V. POSSIBLE EXPERIMENTAL IMPLEMENTATIONS}
\end{center}

In this section, we give a discussion on possible experimental
implementations. First, as discussed above, the condition
\[
2t_2\ll t_3+t_4
\]
[i.e., $\widetilde{\Omega }\gg $ $g^2/\left( m\Delta _c\right) $] needs to
be satisfied, which can be achieved by increasing the Rabi frequency $%
\widetilde{\Omega }$ (i.e., by increasing the intensity of the\textit{\
resonant pulses}). Second, the total operation time
\[
\tau =2t_1+2t_2+t_3+t_4=\pi \Delta _c/g^2+2m\pi \Delta _c/g^2+2\pi /%
\widetilde{\Omega }
\]
should be much shorter than: (i) the energy relaxation time $\gamma _2^{-1}$
of the level $\left| 2\right\rangle $ (note that the level $\left|
3\right\rangle $ is unpopulated during the entire operation), and (ii) the
lifetime of the cavity mode $\kappa ^{-1}=Q/2\pi \nu _c,$ where $Q$ is the
(loaded)\ quality factor of the cavity. These requirements can in principle
be realized, since one can: (i) reduce $\tau $ by increasing the coupling
constant $g$, (ii) increase $\kappa ^{-1}$ by employing a high-$Q$ cavity so
that the cavity dissipation is negligible during the operation, and (iii)
choose qubit systems (e.g., atoms) with long spontaneous decay time of the
level $\left| 2\right\rangle $ or design qubit systems (e.g.,
superconducting devices) so that the energy relaxation time $\gamma _2^{-1}$
of the level $\left| 2\right\rangle $ is sufficiently long.

\begin{figure}[tbp]
\includegraphics[bb=42 301 534 694, width=8.6 cm, clip]{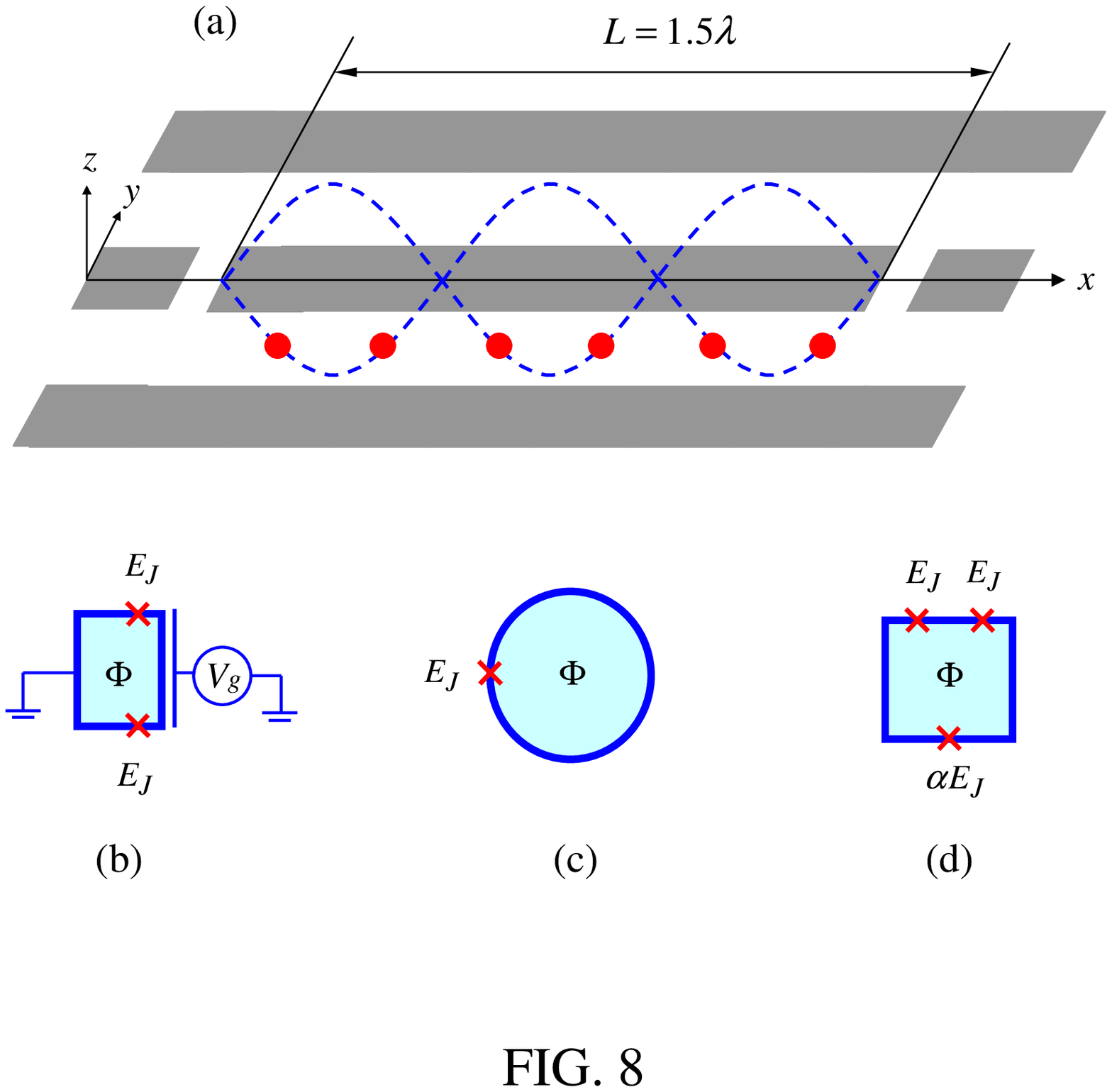} %
\vspace*{-0.08in}
\caption{(color online) (a) Schematic diagram of the setup for six
superconducting qubit systems (red dots) and a (grey) standing-wave
quasi-one-dimensional coplanar waveguide resonator. The two blue curved
lines represent the standing wave magnetic field, which is in the $z$%
-direction. Each qubit system (a red dot) could be a superconducting
charge-qubit system shown in (b), flux-biased phase-qubit system in (c), and
flux-qubit system in (d). The qubit systems are placed at locations where
the magnetic fields are the same to obtain an identical coupling constant $g$
for each qubit system. The superconducting loop of each qubit system, which
is a large square for (b) and (d) while a large circle for (c), is located
in the plane of the resonator between the two lateral ground planes (i.e.,
the $x$-$y$ plane). The external magnetic flux $\Phi$ applied to the
superconducting loop for each qubit system is created by the magnetic field
threading the superconducting loop; $E_{J}$ is the Josephson junction energy
($0.6<\alpha<0.8$) and $V_g$ is the gate voltage; $\lambda$ is the
wavelength of the resonator mode, and $L$ is the length of the resonator.}
\label{fig:8}
\end{figure}

For the sake of definitiveness, let us consider the experimental possibility
of realizing a six-qubit controlled-phase gate in a quantum Fourier
transform (Fig.~9), using six identical superconducting qubit systems
coupled to a resonator [Fig.~8(a)]. Each qubit system could be a
superconducting charge-qubit system [Fig.~8(b)], flux-biased phase-qubit
system [Fig.~8(c)], or flux-qubit system [Fig.~8(d)]. For the present case,
qubit system $1$ acts as a target instead of a control, while qubit systems (%
$2,~3,~4,~5,~6 $) play the role of controls (Fig.~9). Each step of
operations for implementing this gate is the same as that introduced in Sec.
III. As discussed above, to implement this gate, the Rabi frequencies $%
\Omega _2,\Omega _3,\Omega _4,\Omega _5,$ and $\Omega _6$ of the pulses
(respectively applied to qubit systems $2,3,4,5,$ and $6$) need to be set by
$\Omega _{k+1}/\Omega _k=1/\sqrt{2},$ leading to $\chi _{k+1}/\chi _k=1/%
\sqrt{2}$ ($k=2,3,4,5$). By choosing $\Delta _c=10g,\Delta ^{\prime
}=10\Omega _2,\Omega _2\sim 0.9g,\widetilde{\Omega }\sim 10g$ and $m=3,$ we
have: (i) $\delta \sim 10g^2/\Delta _c\sim 10\Omega _2^2/\Delta ^{\prime
}\sim 10\chi _2,$ $\,$(ii) $2t_2=\pi /\widetilde{\Omega }\sim \pi /\left(
5g\right) ;$ and (iii) $t_3+t_4\sim 60\pi /g,$ where $t_3=\left( \pi
/2\right) \left( \delta /\chi _2^2\right) \sim 50\pi /g.$ Therefore, the
condition $2t_2\ll t_3+t_4$ can be well satisfied. Furthermore, for the
parameters chosen here, the total operation time required for the gate
implementation would be $\tau \sim 70\pi /g.$

\begin{figure}[tbp]
\includegraphics[bb=103 348 557 595, width=10.6 cm, clip]{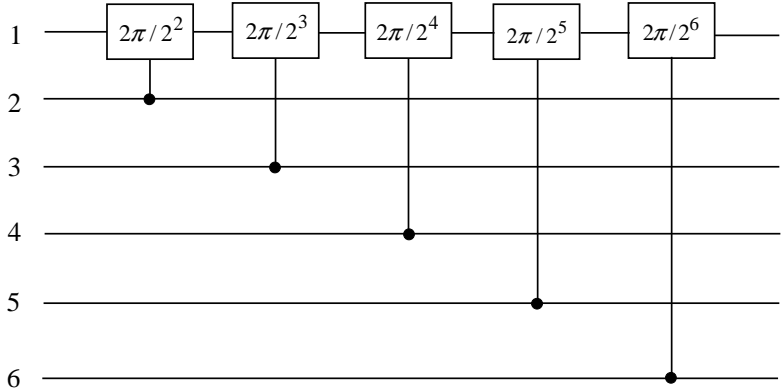} %
\vspace*{-0.08in}
\caption{Circuit for a six-qubit controlled-phase gate for the quantum
Fourier transform. Here, qubits $2,~3,~4,~5,$~and~$6$~(on the filled
circles) play the role of controls acting simultaneously on the target qubit
$1$. The element $2\pi /2^k$ represents a controlled-phase shift $\exp
\left( i2\pi /2^k\right)$, which is performed on the target qubit $1$ when
the control qubit $k$ is in the state $\left| 1\right\rangle $ ($k=2,3,4,5,6$%
). Namely, if the control qubit $k$ is in the state $\left| 1\right\rangle $%
, then a phase shift $2\pi /2^k$ happens to the state $\left| 1\right\rangle
$ of the target qubit $1$, while nothing happens otherwise. }
\label{fig:9}
\end{figure}

As a rough estimate, assume $g/{2\pi }\sim 220$ MHz, which could be reached
for a superconducting qubit system coupled to a one-dimensional
standing-wave CPW (coplanar waveguide) transmission resonator [53]. For the $%
g$ chosen here, we have $\tau \sim 0.16$ $\mu $s, much shorter than $\gamma
_2^{-1}\sim 1$ $\mu $s [2,10]. In addition, consider a resonator with
frequency $\nu _c\sim 3$ GHz (e.g., Ref. [24]) and $Q\sim 10^5$, we have $%
\kappa ^{-1}\sim 5.3$ $\mu $s, which is much longer than the operation time $%
\tau $ here. Note that superconducting coplanar waveguide resonators with a
quality factor $Q>10^6$ have been experimentally demonstrated [54].

How well this gate would work needs to be further investigated for each
particular experimental set-up or implementation. However, we note that this
requires a rather lengthy and complex analysis, which is beyond the scope of
this theoretical work.

\begin{center}
\textbf{VI. COMPARISON WITH PREVIOUS WORK}
\end{center}

Several points related to this work need to be addressed. First, the present
work deals with the realization of multiqubit phase gates for which the
phase shifts on each target qubit are\textit{\ tunable,} as shown above.
Therefore, it is much more general than our previous work [55]. Note that in
Ref.~[55], a method was proposed for the realization of a multiqubit phase
gate which induces a \textit{fixed} phase-shift of $\pi $ on each target
qubit when the control qubit is in the state $\left| 1\right\rangle .$ The
main advantage for the method in [55] is that the cavity mode can be
initially in an \textit{arbitrary} state and thus no preparation of the
initial state for the cavity mode is required. Second, Ref.~[56] showed how
multi-ion GHZ entangled states can be realized based on performing a
multiqubit CNOT gate with one qubit simultaneously controlling $n$ target
qubits. Since their scheme works for a special case (i.e., a CNOT on each
target qubit or a \textit{fixed} phase-shift $\pi $ on each target), it
fails in performing other types of multiqubit phase gates for which the
phase shift on each target qubit is not $\pi $. Third, we note that using
the method presented in [56], a multiqubit CNOT gate or phase gate with a
phase shift $\pi $ on each target qubit (i.e., the first type of multiqubit
gate discussed above) is rather difficult to implement in the other systems
due to different physical mechanisms. Last, the present proposal for
realizing the $n$ successive two-qubit CP gates in QFT is different from the
previous work reported in [57]. As shown above, using the present proposal,
the $n$ successive two-qubit CP gates in QFT can be simultaneously performed
using \textit{one} single-mode cavity only. However, as argued in Ref.~[57],
when using the proposal in [57] to implement $n$ successive two-qubit CP
gates in QFT, $n$ single-mode cavities would be required or a single cavity
with various modes interacting qubit systems (e.g., atoms) needs to be
designed.

\begin{center}
\textbf{VII. CONCLUSION}
\end{center}

We have presented a method to realize a multiqubit \textit{tunable} phase
gate with one qubit simultaneously controlling $n$ target qubits in a cavity
QED. As shown above, the method has the following features: (i) Neither
adjustment of the level spacings of qubit systems nor adjustment of the
cavity mode frequency during the gate operation is needed, thus the
operation is much simplified; (ii) The operation time required for the gate
realization is independent of the number of qubits and thus does not
increase with the number of qubits; (iii) The $n$ two-qubit CP gates
involved can be simultaneously performed, which significantly reduces the
gate operation especially when the number $n$ of qubits is large; (iv) The
phase shift on the state of each target qubit can be \textit{adjusted} by
changing the Rabi frequencies of the pulses applied to the target qubit
systems; and (v) Certain types of significant quantum controlled-phase gates
with one qubit simultaneously controlling multiple target qubits or quantum
controlled-phase gates with multiple control qubits simultaneously acting on
one target qubit (e.g., the multiqubit phase gate in the QFT discussed
above) can be performed by using this proposal.

\begin{center}
\textbf{ACKNOWLEDGMENTS}
\end{center}

CPY thanks Yu-xi Liu for very useful comments, and is also grateful to Sahel
Ashhab, Hou Ian and Jie-Qiao Liao for their help in this work. We
acknowledge partial support from the Laboratory of Physical Sciences,
National Security Agency, Army Research Office, Defense Advanced Research Projects Agency,
Air Force Office of Scientific Research, National Science Foundation
Under Grant.~No.~0726909, DARPA, JSPS-RFBR under Grant No.~09-02-92114,
Grant-in-Aid for Scientific Research (S), MEXT Kakenhi on Quantum
Cybernetics, and FIRST (Funding Program for Innovative R\&D on S\&T). S.B.
Zheng acknowledges support from the National Natural Science Foundation of
China under Grant No. 10674025, the Doctoral Foundation of the Ministry of
Education of China under Grant No. 20070386002, and funds from the State Key
Laboratory Breeding Base of Photocatalysis, Fuzhou University. C.P. Yang also
acknowledges funding support from the National Natural
Science Foundation of China under Grant No. 11074062, the Zhejiang Natural
Science Foundation under Grant No. Y6100098, the funds from Hangzhou
Normal University, and the Open Fund from the SKLPS of ECNU.

\begin{center}
\textbf{APPENDIX: HOW TO HAVE THE LEVEL }$\left| 0\right\rangle $ \textbf{%
NOT TO BE AFFECTED DURING THE OPERATION}
\end{center}

As shown above, three types of interaction are needed for the gate
implementation, which are the system-cavity-pulse resonant Raman coupling,
the system-cavity-pulse off-resonant Raman coupling, and the system-pulse
resonant interaction. For the last type of interaction, since a resonant
pulse is applied, the level $\left| 0\right\rangle $ can be easily made not to be
affected by the pulse, by prior adjustment of level spacings such that the
transition between the level $\left| 0\right\rangle $ and any one of 
other levels is largely detuned (decoupled) from the pulse. In the
following, we will focus on how to have the level $\left| 0\right\rangle $
not to be affected by the pulse and the cavity mode, for the case when the pulse
is off-resonant with the $\left| 1\right\rangle \leftrightarrow \left|
3\right\rangle $ transition and the cavity mode is off-resonant with the $%
\left| 2\right\rangle \leftrightarrow \left| 3\right\rangle $ transition,
i.e., the case which applies to the former two types of interaction.

\begin{figure}[tbp]
\includegraphics[bb=83 268 495 536, width=12.6 cm, clip]{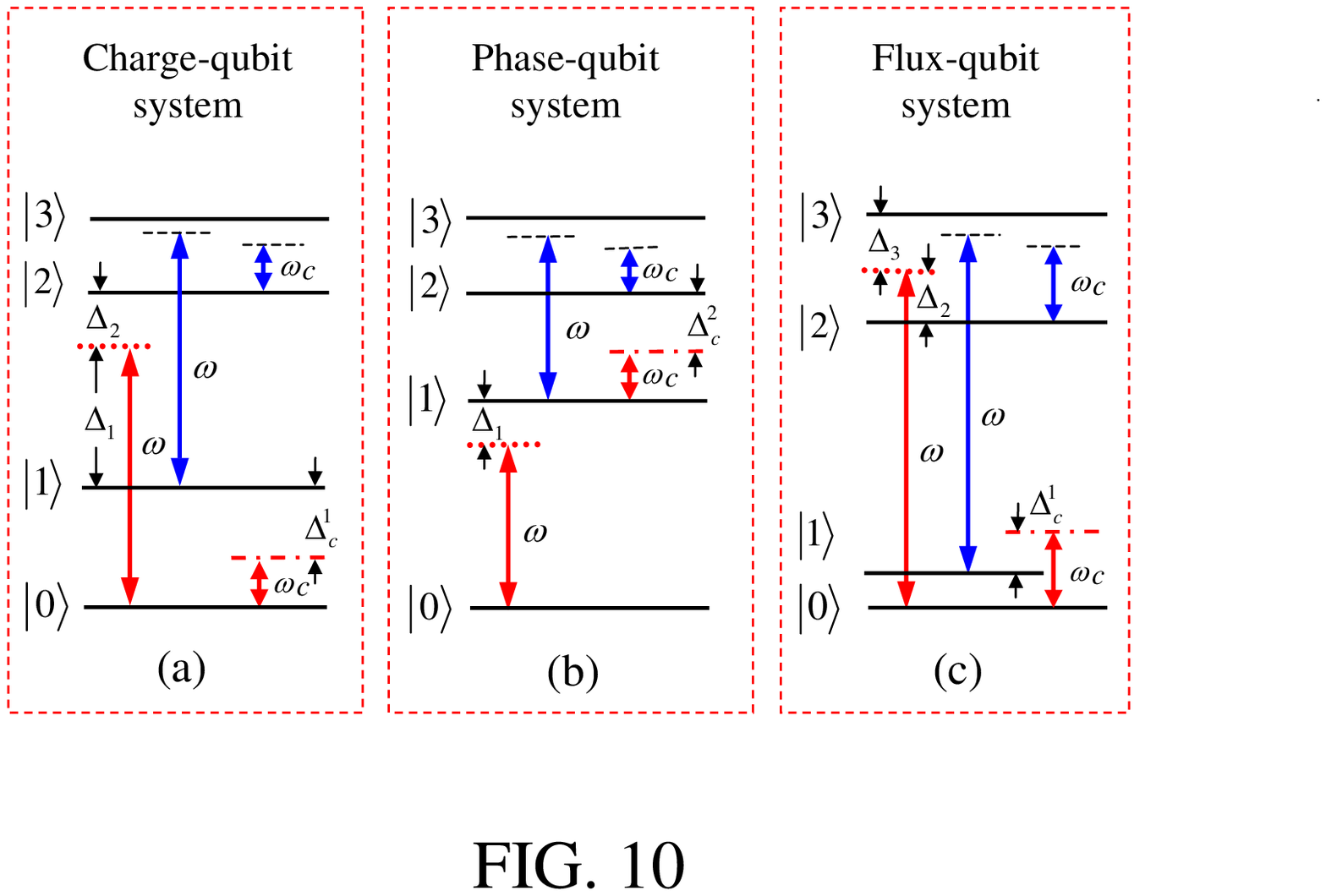} %
\vspace*{-0.08in}
\caption{(Color online) Large detuning of the pulse/the cavity mode with
the transition between the level $\left| 0\right\rangle $ and any one of 
other levels.}
\label{fig:3}
\end{figure}

To begin with, we define $\Delta _1=\left| \left( E_1-E_0\right) /\hbar
-\omega \right| ,$ $\Delta _2=\left| \left( E_2-E_0\right) /\hbar -\omega
\right| ,$ and $\Delta _3=\left| \left( E_3-E_0\right) /\hbar -\omega
\right| $ as the large detuning of the pulse with the $\left| 0\right\rangle
\leftrightarrow \left| 1\right\rangle $ transition, the large detuning of
the pulse with the $\left| 0\right\rangle \leftrightarrow \left|
2\right\rangle $ transition, and the large detuning between the pulse and
the $\left| 0\right\rangle \leftrightarrow \left| 3\right\rangle $
transition, respectively (Fig.~10). Here, $\omega $ is the pulse frequency.
In addition, we define $\Delta _c^1=\left| \left( E_1-E_0\right) /\hbar
-\omega _c\right| $ and $\Delta _c^2=\left| \left( E_2-E_1\right) /\hbar
-\omega _c\right| $ as the large detuning of the cavity mode with the $%
\left| 0\right\rangle \leftrightarrow \left| 1\right\rangle $ transition and
the large detuning of the cavity mode with the $\left| 1\right\rangle
\leftrightarrow \left| 2\right\rangle $ transition, respectively (Fig.~10).

\begin{center}
\textbf{A. Charge-qubit system}
\end{center}

A charge-qubit system has the level structure depicted in Fig.~3(a) [or
Fig.~10(a)]. From Fig.~10(a), one can see that the dotted line falls within
the range between the levels $\left| 1\right\rangle $ and $\left|
2\right\rangle $ in the case when the pulse is off-resonant with the $\left|
1\right\rangle \leftrightarrow \left| 3\right\rangle $ transition. This is
because the level spacing $E_2-E_0$ is larger than the level spacing $E_3-E_1
$, resulting from $E_3-E_2<E_1-E_0$ as mentioned earlier [see the caption of
Fig.~3(a)]. Fig.~10(a) also shows that the dot-dashed line falls within the
range between the levels $\left| 0\right\rangle $ and $\left| 1\right\rangle
$ in the case when the cavity mode is off resonant with the $\left|
1\right\rangle \leftrightarrow \left| 3\right\rangle $ transition. This is
due to $E_3-E_2<E_1-E_0$. In order to have the level $\left| 0\right\rangle $
not to be affected by the pulse and the cavity mode, one will need to adjust the
level spacings of the qubit system before the gate operation to achieve: (i)
the large detunings $\Delta _1$ and $\Delta _2$ [Fig.~10(a)], such that neither $\left|
0\right\rangle \leftrightarrow \left| 1\right\rangle $ transition nor $%
\left| 0\right\rangle \leftrightarrow \left| 2\right\rangle $ transition is
induced by the pulse; and (ii) the large detuning $\Delta _c^1$ [Fig.~10(a)] such that
the cavity mode does not cause the $\left| 0\right\rangle \leftrightarrow
\left| 1\right\rangle $ transition. From Fig.~10(a), it can be seen that
the large detuning regime for
the pulse and the $\left| 0\right\rangle \leftrightarrow \left|
3\right\rangle $ transition is automatically satisfied, and the large detuning of
the cavity mode with the $\left|
0\right\rangle \leftrightarrow \left| 2\right\rangle $ or $\left|
0\right\rangle \leftrightarrow \left| 3\right\rangle $ transition is
also automatically met. Therefore, by prior
adjustment of the level spacings to have the large detunings $\Delta _1,$ $%
\Delta _2$, and $\Delta _c^1,$ the level $\left| 0\right\rangle $ will not
be affected by the pulse and the cavity mode during the operation.

\begin{center}
\textbf{B. Phase-qubit system}
\end{center}

A phase-qubit system has the level structure shown in Fig.~3(b) [or Fig.~
10(b)]. It can be seen from Fig.~10(b) that when the pulse is off resonant
with the $\left| 1\right\rangle \leftrightarrow \left| 3\right\rangle $
transition, the dotted line falls within the range between the two lowest
levels $\left| 0\right\rangle $ and $\left| 1\right\rangle $, which could be
achieved by adjusting the level spacings of the qubit system. In addition,
Fig.~10(b) shows that the dot-dashed line falls within the range between the
levels $\left| 1\right\rangle $ and $\left| 2\right\rangle $, in the case
when the cavity mode is off resonant with the transition between the upper
two levels $\left| 2\right\rangle $ and $\left| 3\right\rangle $. This is
because of $E_2-E_1>E_3-E_2$ [see the caption of Fig.~3(b)]. To have the level $%
\left| 0\right\rangle $ not to be affected by the pulse and the cavity mode, one
will need to adjust the level spacings of the qubit system before the gate
operation to obtain: (i) the large detuning $\Delta _1$ [Fig.~10(b)] such that the $%
\left| 0\right\rangle \leftrightarrow \left| 1\right\rangle $ transition
induced by the pulse is avoided; and (ii) the large detuning $\Delta _c^2$ which ensures
the large detuning between the cavity mode and the $\left|
0\right\rangle \leftrightarrow \left| 1\right\rangle $ transition 
due to $E_1-E_0>E_2-E_1$ [Fig.~10(b)].
In addition, Fig.~10(b) shows that the large detuning regime for the pulse and the $\left|
0\right\rangle \leftrightarrow \left| 2\right\rangle $ or $\left|
0\right\rangle \leftrightarrow \left| 3\right\rangle $ transition is
automatically met, and the large detuning of the
cavity mode with the $\left| 0\right\rangle \leftrightarrow \left|
2\right\rangle $ or $\left| 0\right\rangle \leftrightarrow \left|
3\right\rangle $ transition is automatically satisfied.
Thus, as long as the large detunings $\Delta _1$ and $\Delta _c^2$
are met by prior adjustment of the level spacings, the level $\left|
0\right\rangle $ will not be affected by the pulse and the cavity mode during
the operation.

\begin{center}
\textbf{C. Flux-qubit system}
\end{center}

A flux-qubit system has the level structure in Fig.~3(c) [or Fig.~10(c)].
Fig.~10(c) shows that the dotted line falls within the range between the
levels $\left| 2\right\rangle $ and $\left| 3\right\rangle $ in the case
when the pulse is off-resonant with the $\left| 1\right\rangle
\leftrightarrow \left| 3\right\rangle $ transition. This is because the
level spacing $E_2-E_0$ is smaller than the level spacing $E_3-E_1$ due to 
$E_1-E_0<E_3-E_2$ [see the caption of Fig.~3(c)]. Fig.~10(c)
also shows that the dot-dashed line is above
the level $\left| 1\right\rangle ,$ which could be reached by adjusting the
level spacings of the qubit system. In order to have the level $\left|
0\right\rangle $ not to be affected by the pulse and the cavity mode, one will
need to adjust the level spacings of the qubit system to achieve: (i) the
large detunings $\Delta _2$ and $\Delta _3$ [Fig.~10(c)], such that neither $\left|
0\right\rangle \leftrightarrow \left| 2\right\rangle $ transition nor $%
\left| 0\right\rangle \leftrightarrow \left| 3\right\rangle $ transition is
induced by the pulse; and (ii) the large detuning $\Delta _c^1$ [Fig.~10(c)], in
order to avoid the
$\left| 0\right\rangle \leftrightarrow \left| 1\right\rangle $ transition
induced by the cavity mode. From Fig.~10(c), it can be seen that
the large detuning regime for the
pulse and the $\left| 0\right\rangle \leftrightarrow \left| 1\right\rangle $
transition is automatically satisfied, and the large detuning of the cavity
mode with the $\left| 0\right\rangle \leftrightarrow \left| 2\right\rangle $
or $\left| 0\right\rangle \leftrightarrow \left| 3\right\rangle $ transition
is automatically met. Hence, as long as the large detunings $\Delta _2,$ $%
\Delta _3,$ and  $\Delta _c^1$ are satisfied by prior adjustment of the level
spacings, the level $\left| 0\right\rangle $ will not be affected during the
operation. We here should mention that because the dotted line falls within
the range between the upper two levels $\left| 2\right\rangle $ and $\left|
3\right\rangle $ [Fig.~10(c)], a more careful adjustment of the level
spacings would be needed to obtain the large detunings $\Delta _1$ and $%
\Delta _2$ for the flux-qubit system, when compared with the charge-qubit
system or the phase-qubit system.

\end{document}